\documentclass[]{aa}
\usepackage[varg]{txfonts}
\usepackage{graphicx}
\usepackage{booktabs}
\usepackage{amsmath}
\usepackage{colortbl}
\usepackage{subfig}
\usepackage{rotating}
\usepackage{amssymb}
\usepackage{latexsym}
\usepackage{ifthen}
\usepackage{enumitem}
\usepackage{color}

\begin{document}

\title{LBT/ARGOS  adaptive optics observations of  z$\sim 2$ lensed galaxies\thanks{The LBT is an international collaboration among institutions in the United States, Italy and Germany. LBT Corporation partners are: The University of Arizona on behalf of the Arizona university system; Istituto Nazionale di Astrofisica, Italy; LBT Beteiligungsgesellschaft, Germany, representing the Max-Planck Society, the Astrophysical Institute Potsdam, and Heidelberg University; The Ohio State University, and The Research Corporation, on behalf of The University of Notre Dame, University of Minnesota and University of Virginia.}} 
\titlerunning{LBT/ARGOS  AO observations of z$\sim 2$ lensed galaxies} 
\author{M. Perna
                \inst{\ref{i1}}\thanks{E-mail: perna@arcetri.inaf.if}
                \and 
         M. Curti
                \inst{\ref{i1},\ref{i2},\ref{i3},\ref{i4}}
                \and
            G. Cresci
                \inst{\ref{i1}}        
                \and
        F. Mannucci
                \inst{\ref{i1}} 
                \and
         S. Rabien
                \inst{\ref{i5}}
                \and
         C. Grillo
                \inst{\ref{i8},\ref{i9}}
                \and 
         S. Belli
                \inst{\ref{i5}}
                \and
        M. Bonaglia 
                \inst{\ref{i1}}
                \and
         L. Busoni
                \inst{\ref{i1}}
                \and
      A. Contursi
                \inst{\ref{i5}}
               \and
         S. Esposito
                \inst{\ref{i1}}
                \and
         I. Georgiev
                \inst{\ref{i7}}
                \and
         D. Lutz
                \inst{\ref{i5}}
                \and 
         G. Orban de Xivry
                \inst{\ref{i5},\ref{i10}}
                \and 
         S. Zibetti
                \inst{\ref{i1}}
                \and
         W. Gaessler
                \inst{\ref{i7}}
                \and 
         T. Mazzoni
                \inst{\ref{i1}}
                \and 
         J. Borelli
                \inst{\ref{i7}}
                \and 
         M. Rosensteiner
                \inst{\ref{i5}}
                \and 
         J. Ziegleder
                \inst{\ref{i5}}
                \and 
         P. Buschkamp
                \inst{\ref{i5}}
                \and 
         G. Rahmer
                \inst{\ref{i11}}
                \and 
         M. Kulas 
                \inst{\ref{i7}}
                \and 
         D. Peter
                \inst{\ref{i7}}
                \and 
         W. Raab
                \inst{\ref{i5}}
                \and 
         M. Deysenroth
                \inst{\ref{i5}}
                \and 
         H. Gemperlein
               \inst{\ref{i5}}
}

\institute{
        INAF - Osservatorio Astrofisico di Arcetri, Largo Enrico Fermi 5, I-50125 Firenze, Italy\label{i1}
        \and
        Dipartimento di Fisica e Astronomia, Universit\`a di Firenze, via G. Sansone 1, I-50019 Sesto Fiorentino (Firenze), Italy\label{i2}
        \and
        Cavendish Laboratory, University of Cambridge, 19 J. J. Thomson Ave., Cambridge CB3 0HE, UK\label{i3}
        \and
        Kavli Institute for Cosmology, University of Cambridge, Madingley Road, Cambridge CB3 0HA, UK\label{i4}
        \and
        Max-Planck-Institut f\"ur Extraterrestrische Physik (MPE), Giessenbachstr. 1, D-85748 Garching, Germany\label{i5}
        \and
        Max-Planck Instiut f\"ur Astronomie, K\"onigstuhl 17, D-69117 Heidelberg, Germany\label{i7}
        \and
        Dipartimento di Fisica, Universit\`a degli Studi di Milano, via Celoria 16, I-20133 Milano, Italy\label{i8}
        \and
        Dark Cosmology Centre, Niels Bohr Institute, University of Copenhagen, Juliane Maries Vej 30, DK-2100 Copenhagen, Denmark\label{i9}
        \and
        Space sciences, Technologies and Astrophysics Research Institute, Universit\'e de Li\`ege, All\'ee du Six Ao\^ut 17, 4000 Li\`ege, Belgium\label{i10}
        \and 
        Large Binocular Telescope Observatory, Tucson, Arizona, USA\label{i11}
}

\date{Received 2 November 1992 / Accepted 7 January 1993}

\abstract {} 
{
Gravitationally lensed systems allow a detailed view of galaxies at high redshift. High spatial- and spectral-resolution measurements of arc-like structures can offer unique constraints on the physical and dynamical properties of high-z systems.
} 
{
We present  near-infrared spectra centred on the gravitational arcs of six known z $\sim 2$ lensed star-forming galaxies of stellar masses of 10$^{9-11}$ M$_{\odot}$ and star formation rate (SFR) in the range between 10 and 400 M$_{\odot}$/yr. Ground layer adaptive optics (AO)-assisted observations are obtained at the Large Binocular Telescope (LBT) with the LUCI spectrographs during the commissioning of the ARGOS facility. We used MOS masks with  curved slits to follow the extended arched structures and study the diagnostic emission lines. LBT observations are used to demonstrate the spectroscopic capabilities of ARGOS.  
} 
{
Combining spatially resolved kinematic properties across the arc-like morphologies, emission line diagnostics and archival information, we distinguish between  merging and rotationally supported systems, and reveal the possible presence of ejected gas. For galaxies that have evidence for outflows, we derive outflow energetics and mass-loading factors compatible with those observed for stellar winds in local and high-z galaxies. We also use flux ratio diagnostics to derive gas-phase metallicities. The low signal-to-noise ratio in the faint H$\beta$ and nitrogen lines allows us to derive an upper limit of  $\approx 0.15$ dex for the spatial variations in metallicity along the slit for the lensed galaxy J1038.
} 
{
Analysed near-infrared spectra presented here represent the first scientific demonstration of performing AO-assisted multi-object spectroscopy with narrow curved-shape slits.
The increased angular and spectral resolution, combined with the binocular operation mode with the 8.4-m-wide eyes of LBT, will allow the characterisation of kinematic and chemical properties of a large sample of galaxies at high-z in the near future. 
} 

\maketitle
\titlerunning{X-ray loudness and outflows} 

\section[Introduction]{Introduction}

\begin{table*}[h]
\centering
\scriptsize
\caption{Galaxy properties and log of observations.}
\label{tablog}
\begin{tabular}{lccccc|ccccccc}
Name & z & Coordinates & $\mu$& log($M_*$) & SFR& Dates& Band & width & TOT & seeing &FWHM&resolution\\
\tiny{  (1)} & \tiny{  (2) } &\tiny{  (3) } &\tiny{   (4) } &\tiny{  (5) }& \tiny{  (6) } &\tiny{  (7) } &\tiny{  (8) } &\tiny{  (9) }& \tiny{  (10) } &\tiny{  (11) } &\tiny{  (12) }  &\tiny{  (13) }\\

\toprule
  J0143 (CSWA116)& 1.502 & 01:43:50 +16:07:41 &11& $10.1_{-0.1}^{+0.1}$& $75_{-4}^{+4}$           & 10/17    & J &     0.7$''$   &   35m   &   1.30$''$     &    $0.50''$          &     1.3 kpc        \\
          &          &                                  & &&&                   & H &   0.7$''$    & 35m     & 1.10$''$& $0.42''$ & 1.1 kpc     \\
\hline
J1343 (CSWA28)& 2.092 & 13:43:36 +41:54:29 & 9& $10.0_{-0.3}^{+0.3}$&12$_{-11}^{+11}$& 05/16    & K & 0.5$''$ & 35m    & 0.97$''$ &0.41$''$ & 0.7 kpc \\
\hline
J1038 & 2.197 & 10:38:42 +48:49:19 & 8 & $9.1_{-0.1}^{+0.2}$ &40$_{-2}^{+2}$&03/17 & H & 0.3$''$ & 260m  & 1.00$''$&0.52$''$& 1.0 kpc\\
(Cheshire cat arc) &           &                                 & &&&               & K &        & 227m &  1.00$''$& 0.43$''$& 0.8 kpc\\
\hline
J1958 (CSWA128)& 2.225 & 19:58:35 +59:50:59 & 10&$10.7_{-0.4}^{+0.4}$&250$_{-142}^{+71}$& 05/17    & H & 0.3$''$ & 30m     & 1.18$''$&0.50$''$& 0.6 kpc\\
          &          &                                  && &&                   & K &             & 27m     & 1.03$''$& 0.32$''$ & 0.4  kpc     \\
\hline
J1110 (CSWA104)& 2.480 & 11:10:16 +64:59:17 & 9&$9.5_{-0.4}^{+0.4}$&15$_{-4}^{+15}$&05/16 & K & 0.5$''$ & 80m  & 0.98$''$ &0.41$''$ & 0.9 kpc\\
\hline
J0022 (8 o'clock)& 2.736 & 00:22:41 +14:31:14 & 6&10.5$_{-0.6}^{0.5}$ &$390_{-20}^{+20}$& 10/16;10/17 & (K) & 0.3$''$ & 242m  &1.04$''$ &0.43$''$ & 0.9 kpc\\
\toprule  
\end{tabular}
\tablefoot{Column (1): lensed galaxy names as reported in the paper, with other common names in parentheses. Column (2): spectroscopic redshifts from the literature. Column (3): Coordinates of the lensed sources (RA and Dec). Columns (4) and (5): magnification factors and stellar masses, with nominal correction from a Chabrier to a Salpeter IMF when required, from \citet[][J1038]{Jones2013}, \citet[][J0022]{Shirazi2014}, \citet[][J1343]{Saintonge2013}, \citet[][J1110]{Johnson2017}, \citet[][J0143]{Kostrzewa2014}. For J1038, J0022 and J1110 we specifically refer to the A2 knot measurements, as reported in the aforementioned papers; all other values do not refer to individual images, but to the lensed galaxy. Stellar mass of J1958 from this work (see Sect. \ref{gasmetallicity}).  All stellar masses are corrected for the lensing magnification factors reported in Column (4).  Column (6): SFR from the H$\alpha$ flux corrected for extinction and lensing magnification using Kennicutt 1998 relation (J1958 and J1343 from \citealt{Leethochawalit2016}; J0022 from \citealt{Shirazi2014}, using the H$\beta$ flux).  Lensing-corrected SFR of  J1110 from \cite{Johnson2017} and obtained by SED fitting analysis, nominally corrected from a Chabrier to a Salpeter IMF. Column (7): dates (MM/YY). Column (8): LUCI filters; for J0022, we used a clear filter to detect H$\alpha$ line redshifted at $\sim 2.45\mu$m. Column (9): curved slit width. 
Column (10): Time on target. Column (11): seeing as measured by the DIMM station. Column (12): spatial profile FWHM obtained integrating the  reference star flux within a 100 px interval along the X axis. 
Column (13): estimate of the spatial resolution in the source plane. For J1038, J0022 and J1110 we refer to the A2 knots, using the magnifications derived by \cite{Jones2013}, \cite{Shirazi2014} and \cite{Johnson2017}, respectively. 
}
\end{table*}

The history of baryonic assembly in galaxies is governed by complex interactions between different astrophysical processes. The accretion of cool gas through cosmological infall of pristine and recycled material regulates both the star formation and the growth of the central supermassive black hole (SMBH) in the host galaxy (e.g. \citealt{Lilly2013}). Galactic-scale outflows associated with stellar activity and SMBH accretion, depositing energy and momentum to the surrounding medium, are expected to affect the physical and dynamical conditions of infalling gas and the formation of new stars, preventing massive galaxies from overgrowing, and producing the red colours of ellipticals (see \citealt{Somerville2015} for a recent review).

Most star-forming galaxies (SFGs) and active galactic nuclei (AGN) show evidence of powerful, galaxy-wide outflows. Spatially resolved optical, infrared (IR) and millimetre spectroscopic studies are revealing the extension, the morphology, and the energetics related to the ejected multiphase material, both in the local Universe (e.g. \citealt{Arribas2014,Cresci2015b,Cicone2014,Contursi2013,Feruglio2015,Harrison2014,Rupke2013,Sturm2011}) and at high redshift (e.g. \citealt{Brusa2016,Schreiber2014,Genzel2014,Harrison2012,Harrison2016,Newman2012,Perna2015a,Perna2015b}). In order to understand the physical processes responsible for the outflows and the long-term effects they may have on the global baryon cycle, host galaxy and outflow properties have been studied looking for causal relationships. 

In AGN-dominated systems, outflow energetics seem to be related to the AGN power: faster and more powerful winds are preferentially found in more luminous, massive, and fast accreting nuclei (e.g. \citealt{Carniani2015,Fiore2017,Harrison2016,Perna2017a,Woo2016,Zakamska2014}). AGN-driven outflow properties have also been related to star formation rate (SFR) in the host galaxy in order to test the impact of these winds on the state of star forming gaseous clouds, but contrasting results have been found ({e.g. \citealt{Balmaverde2016,Lanzuisi2017,Mullaney2015,Wylezalek2016}).

Star-formation (SF)-driven winds have also been investigated and several authors have found that the incidence of outflows increases with the SFR (e.g. \citealt{Cicone2016,Chen2010,Rubin2014}).
In SF-dominated systems, the effects of outflows can also be probed studying the relationships between SFR, stellar mass, and gas-phase metallicity (\citealt{Mannucci2010,Tremonti2004}), because of the interconnection between accretion/ejection of material to/into the interstellar medium (ISM) and the physical conditions regulating the formation of new stars (e.g. \citealt{Dave2011,Lilly2013}).
In particular, the spatial distribution of the heavy elements within a galaxy, keeping track of the effects of outflows, can constrain the baryonic assembly history  (e.g. \citealt{Angles2014,Gibson2013}).

Lensed galaxies and quasars offer a wealth of information to study galaxy evolution at redshifts in the range $1.5 - 3$, that is, at the peak of star formation and AGN activity: magnification effect increases both the apparent size on the sky and the total flux of lensed sources by up to factors
of tens. Ionised gas emission lines can therefore be used to characterise the internal kinematics and metallicity gradients on scales as small as 100 parsec, especially when exploiting the correction provided by adaptive optics (AO; e.g. \citealt{Jones2010,Jones2013,Leethochawalit2016,Shirazi2014,Stark2008}). In particular, the study of gravitationally lensed systems represents an ideal strategy to characterise the less massive systems at z $\gtrsim$ 2.

In this paper, we present kinematic and chemical analyses of the first
combined LUCI and ARGOS observations of six bright lensed galaxies at
redshifts of z $=1.5-2.7$. LUCI is an IR multi-mode instrument
installed at the Large Binocular Telescope (LBT) that can be assisted by the wide-field ``seeing enhancer'' ARGOS facility (see Sect. \ref{Observations}); a detailed description of the general performances provided by ARGOS, both in terms of spectroscopic and imaging capabilities, is presented in Rabien et al., in prep. 
The targets were selected on the basis of their redshifts (in order to ensure the presence of the emission lines of interest in the proper wavelength range), lensing magnification, and the presence of a nearby suitable stars (R $<16$) for tip-tilt correction. Their visibility within the scheduled ARGOS commissioning runs was also taken into account. 

The lensed galaxies are characterised by SFR in the range between 10 and 400 M$_{\odot}$/yr, and stellar masses $9.1<$ log(M$_*$/M$_{\odot}$) $<10.7$. 
All but one galaxy (J1343+4154; J1343 hereinafter) are above the main sequence in SFR-M$_*$ plane at z $\sim 2$ (e.g. \citealt{Whitaker2012}; see Table \ref{tablog}).
We characterise the physical properties of the lensed galaxies by
combining the results we obtained by high-spectral- and
angular-resolution ARGOS data analysis with the available
multiwavelength information, including recent lens modelling presented
by \citet{Jones2013}, \citet{Leethochawalit2016}, \cite{Shirazi2014},
and \citet{Johnson2017}.

\begin{figure*}[th]
\centering
\includegraphics[width=4.8cm,trim=1 0 0 0,clip]{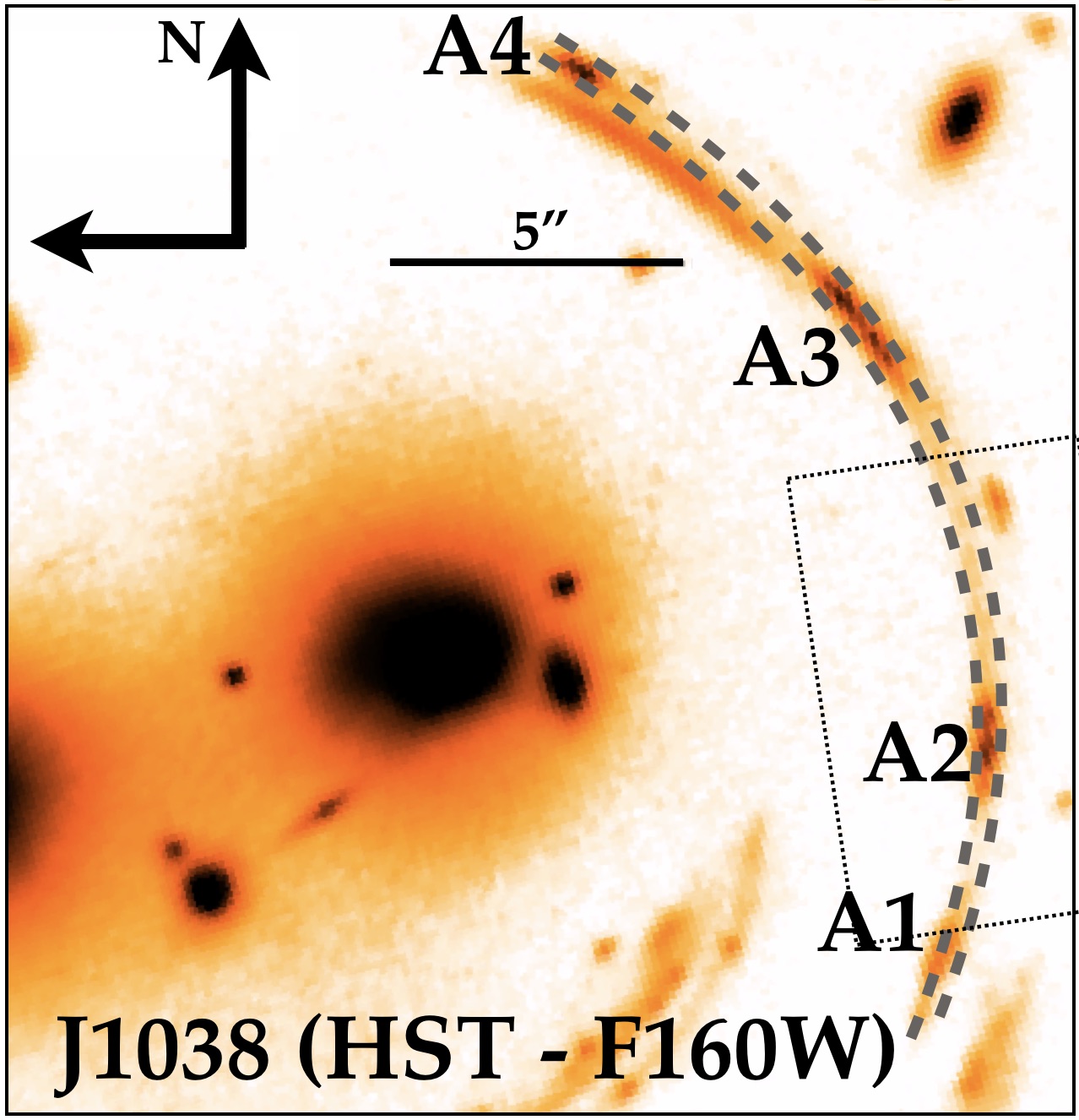}
\hspace{0.01cm}
\includegraphics[width=4.8cm,trim=0 0 0 0,clip]{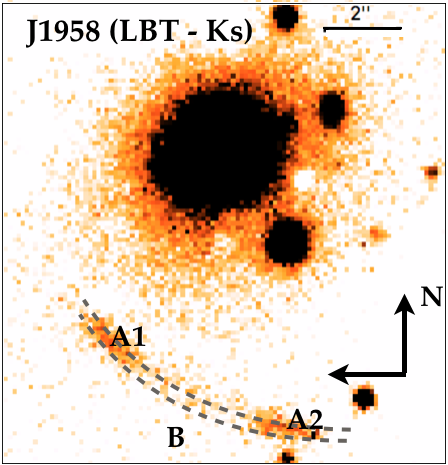}
\hspace{0.01cm}
\includegraphics[width=4.8cm,trim=0 0 0 0,clip]{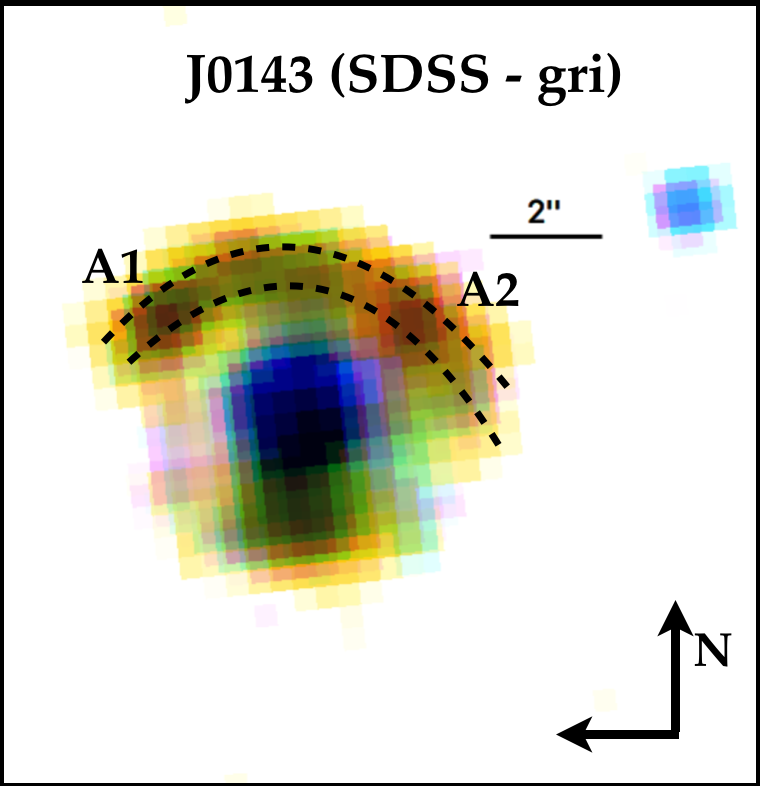}
\hspace{0.01cm}
\includegraphics[width=4.8cm,trim=0 0 0 0,clip]{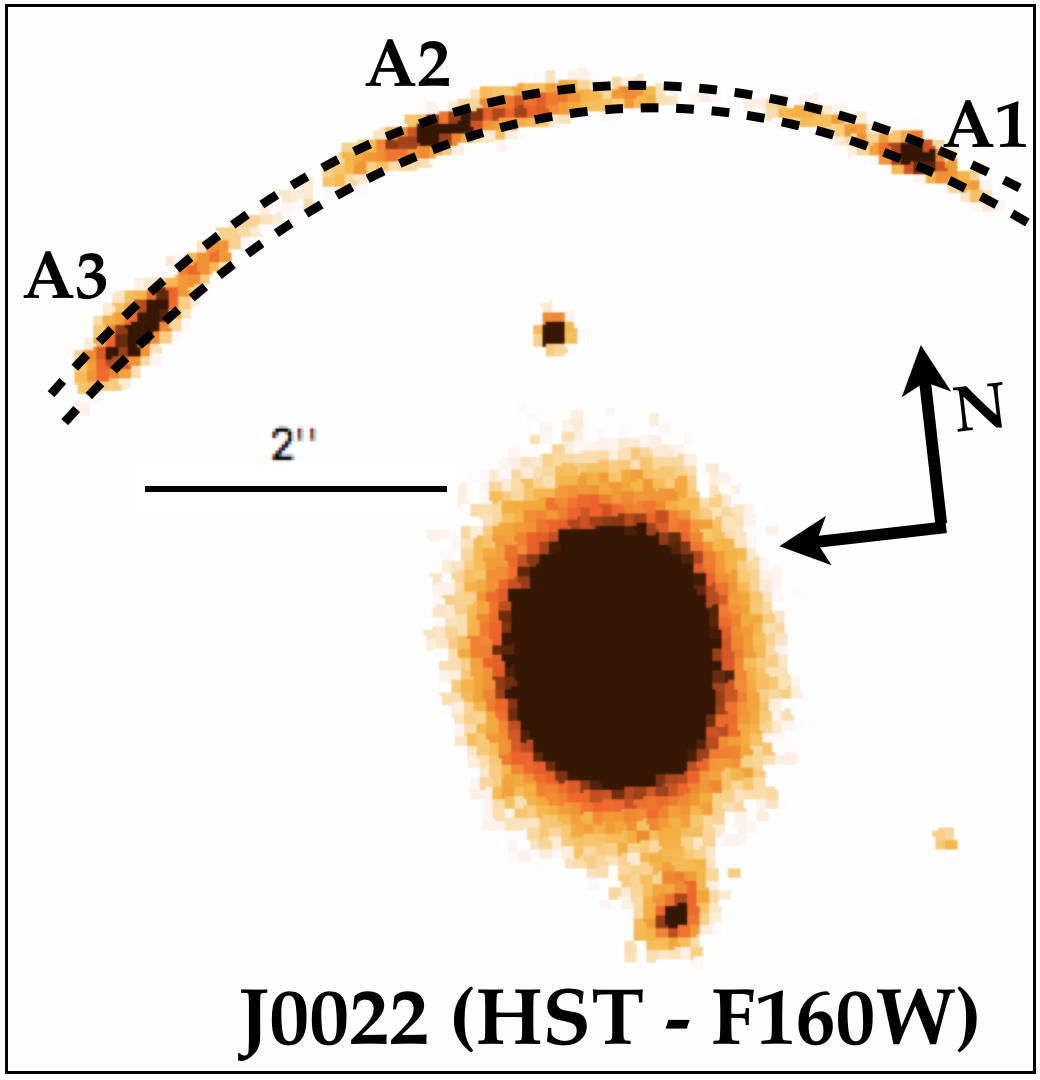}
\hspace{0.01cm}
\includegraphics[width=4.8cm,trim=0 0 0 0,clip]{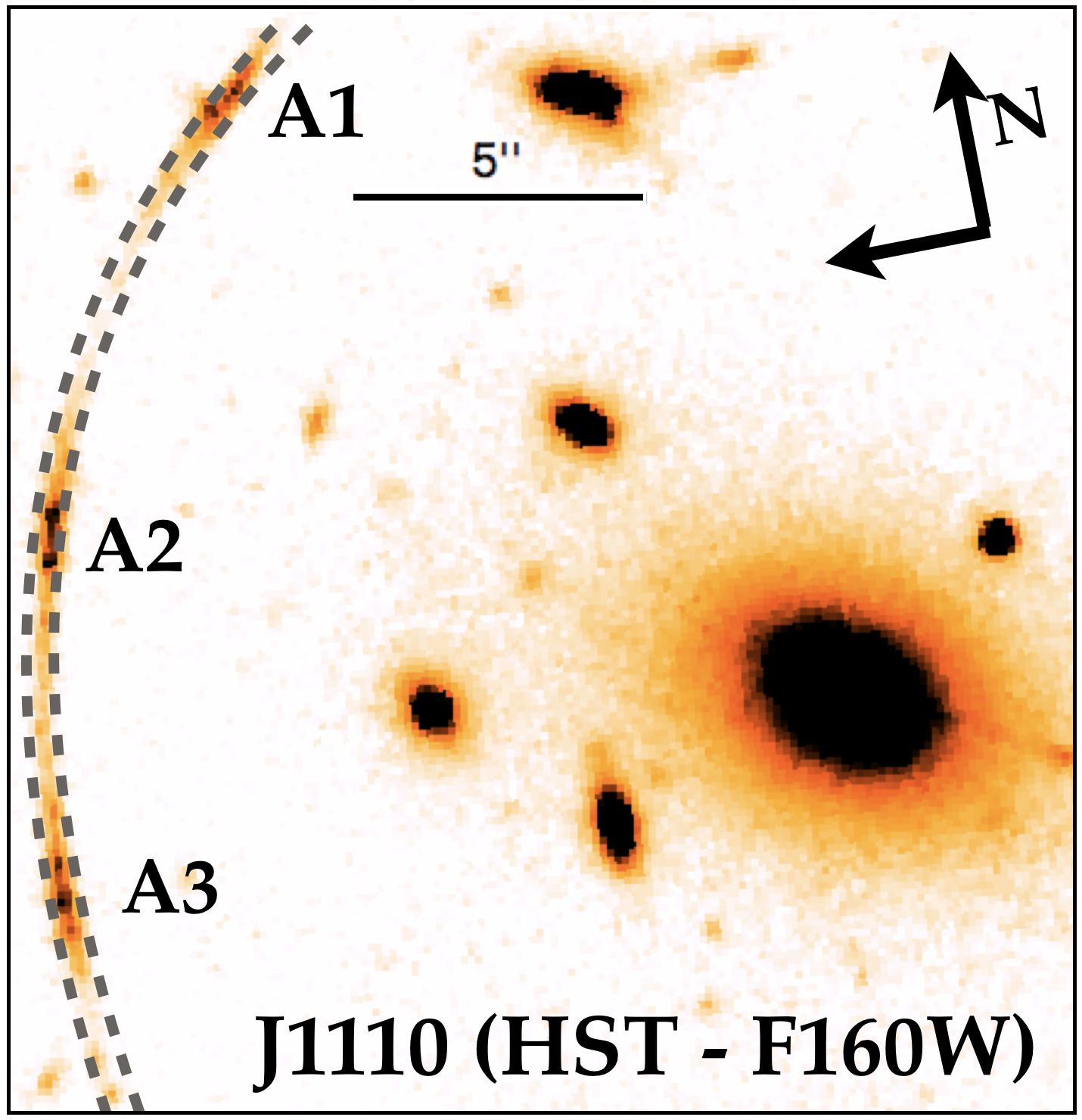}
\hspace{0.01cm}
\includegraphics[width=4.8cm,trim=0 0 0 0,clip]{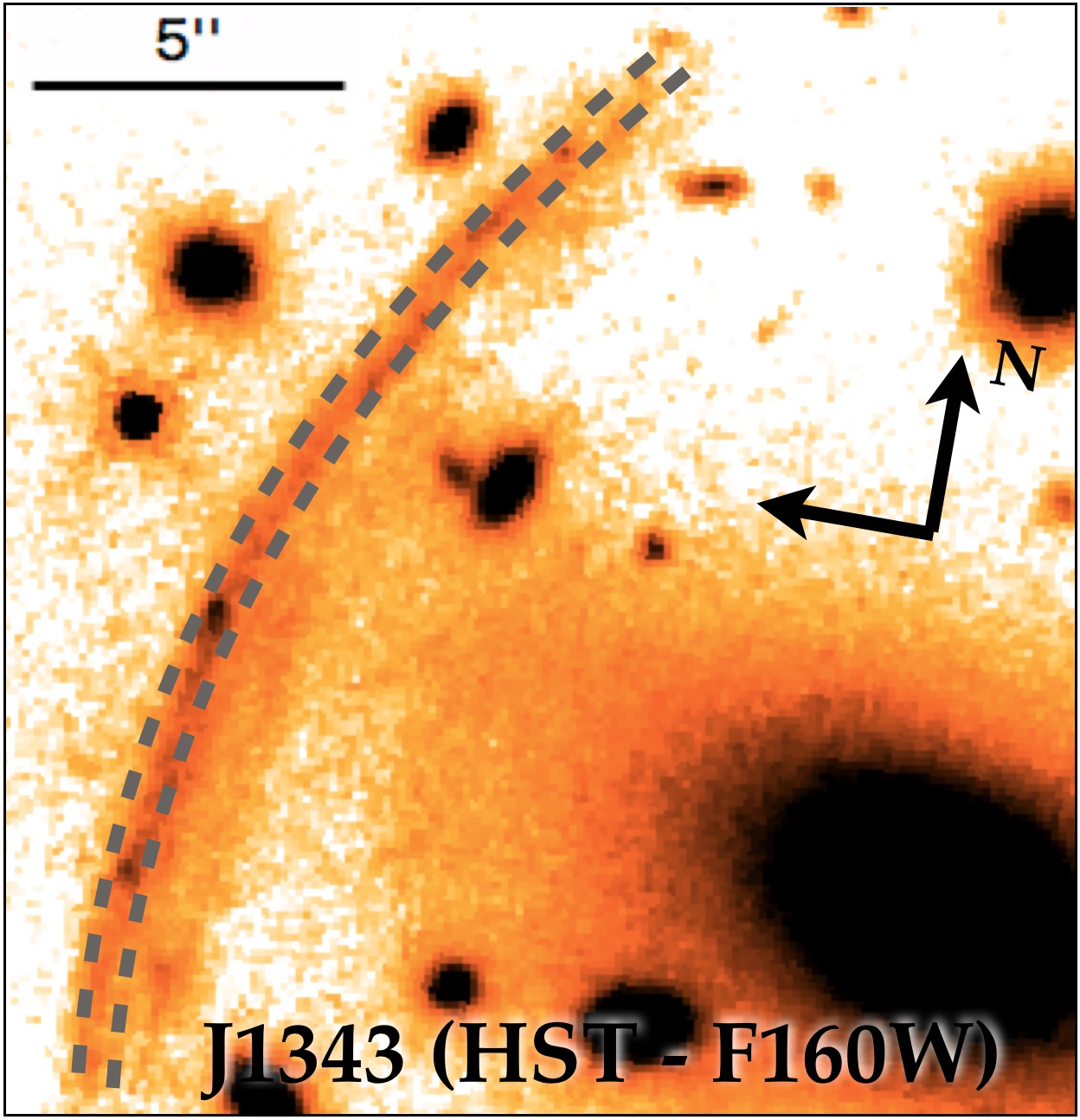}
\caption{\small Infrared HST and LBT/ARGOS images and optical SDSS three-colour cut-out showing the sample of lensed galaxies studied in this paper. Scale bars and target names are labelled in each panel. We identified the lensed galaxy components according to the definitions proposed in this work. For each cut-out, we show the curved slit we constructed for observing the targets (dashed curves). For J1038, we also show with a dotted box the field of view covered by previous observations obtained by \citet{Jones2013} with the IFU instrument OSIRIS. 
}
\label{images}
\end{figure*}

Integral field and/or long-slit spectroscopy of our targets have already been presented in the literature, although most of these observations were not covering the full arcs. We show that our ARGOS/LUCI observations provide higher signal-to-noise-ratio (S/N) spectra, and, thanks to the curved shape of the slits, allow us to cover the total extension of the lensed systems.

The paper is organised as follows: Sections 2 and 3 presents the LBT observations and spectroscopic data reduction. Section 4 presents the data analysis and the spatially resolved spectroscopic results for individual targets. In Section 5, we discuss the possible effects of non-uniform slit illumination on our kinematic analysis. In Section 6, we derive flux ratio diagnostics to characterise the ionisation source of emitting gas. In Section 7, we derive the outflow properties for the sources for which we collected evidence of SF-driven winds in the host galaxy. In Section 8, we study the gas-phase metallicity conditions. Finally, Section 9 summarises our conclusions.
A flat universe model with a Hubble constant of $H_0=$ 72 km s$^{-1}$ Mpc$^{-1}$, $\Omega_M$= 0.27 and $\Omega_{\Lambda}$= 0.73 is adopted. A Salpeter initial mass function (IMF) is adopted to derive stellar masses and SFRs.

\section{Observations}\label{Observations}
The near-infrared (NIR) spectroscopic data presented in this paper were obtained in single and binocular mode with the two NIR multi-object spectrographs (LUCIs), at the LBT on Mount Graham, Arizona, USA.  
The lensed galaxies were observed during the commissioning of the Advanced Rayleigh Guided Ground Layer Adaptive Optics (ARGOS) facility (\citealt{Rabien2010,Busoni2015}). ARGOS is a multi-laser guide-stars ground layer adaptive optics system capable of correcting for the ground-layer turbulence above the two 8.4m LBT mirrors and  provides an effective seeing improvement of a factor of $\sim2$ over a  field of view of $4 \times 4$ arcmins ({\citealt{Orban2015,Orban2016}; Rabien et al., in prep). ARGOS is designed for operation with both LUCI multi-object spectrographs and NIR imagers. 
With the binocular operation mode, it is possible to obtain simultaneous acquisitions with identical or independent instrument configurations for the two LBT mirrors (e.g. taking, for a given target, the spectra in two bands), significantly reducing the time needed for an observing program.

The targets were acquired with dithered 150s observations in an AB sequence. Three or more reference stars are selected for mask alignment and to derive the exact dithering pattern 
during the sequence of acquisitions. 
For four out of six lensed galaxies, curved-slits of  $0.3-0.5''$ in width matching the arc's geometry on the sky have been constructed on the basis of available high-resolution HST images using the Luci Mask Simulator software\footnote{The Luci Mask Simulator (LMS) software is available for download at  \url{http://abell.as.arizona.edu/~lbtsci/Instruments/LUCIFER/lucifer.html}. The LMS version for masks with curved-slits, not officially released, can be made available upon request.}. The other two lensed galaxies in our sample are not covered by HST images. The slit of J1958+5950 (J1958 hereinafter) has therefore been constructed taking advantage of LUCI/ARGOS images taken during the commission time. For J0143+1607 (J0143 hereinafter), we could not obtain high-resolution pre-images for the mask preparation. Therefore, we used the r-band SDSS image to construct the MOS mask, preferring a wider curved slit (0.7$''$) to follow the extended arc and minimise the flux losses. We also note here that those reported in this paper are the first AO-assisted observations of this system. 
In Fig. \ref{images}, we reported the cut-outs for each lens galaxy, with the constructed curved-slits.

All but one target were observed in one or two NIR band filters, depending on the spectroscopic redshift. The H$\alpha$ line of J0022+1431 (J0022 hereinafter), a lensed system at z $\sim 2.7$, is redshifted at $\sim 2.45 \mu$m; in this wavelength region, the K filter transmission is very low, and the line is expected where the thermal continuum contribution is significant. We therefore used an open (``clear'') position in the filter wheel, allowing the spectrograph to capture the entire light spectrum.
  For all but one target, we used the combination of the G210 grating and the N3.75 camera for nominal resolutions of R $\sim 5000$ and $5900$ for H and K bands, respectively, and a pixel scale of $\sim 0.12''/$pixel. G150 grating, with a nominal R of $\sim 4150$ for the K-band, was used for J1343\footnote{All spectral resolutions are given for a 0.5$''$ slit (see Tab. 5 of \citealt{Rothberg2016}).}. Table \ref{tablog} reports the most important properties of the targets and the observations log.

\begin{figure*}[t]
\centering
\includegraphics[width=18.cm,trim=0 0 0 0,clip]{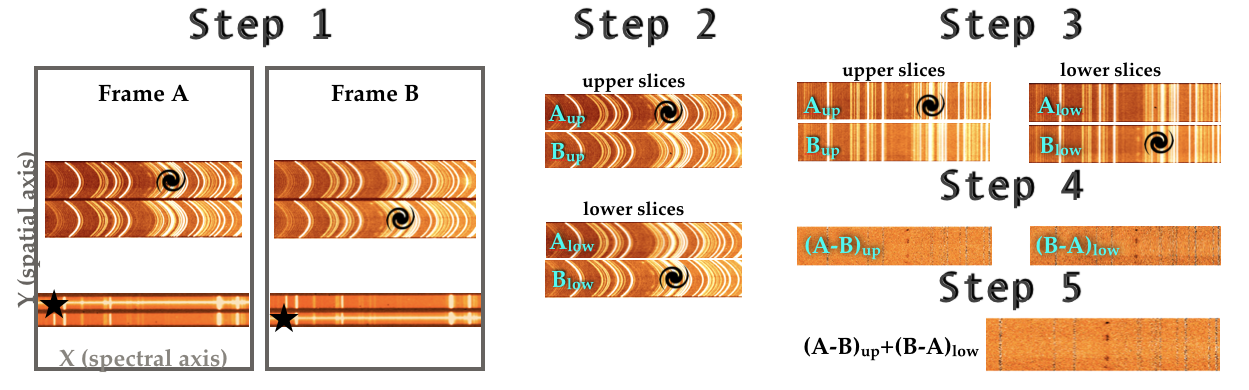}
\caption{\small Sketch to illustrate some of the steps of the data reduction (see Sect. \ref{datareductionsect} for the complete data-reduction procedure used in this work). For clarity, the diagram shows a set of only two subsequent frames. The first two panels show a sketch of a 150s exposure through a MOS mask designed to observe a science target and a reference star with two dither positions. The panels display the 2D spectra obtained from the two curved slitlets (in the upper side), dedicated to the science target, and two long slits (in the lower part), to allocate the reference star spectrum. The extended structures observed in the upper spectra, due to sky emission lines, highlight the strong distortion caused by the curved slitlets.  In the first panel (frame A), the target and the reference star are in the relative upper positions: J1038+4849 H$\alpha$ emission line blobs can be distinguished in the figure (in the vicinity of the spiral symbol), as well as the stellar continuum (next to the star symbol). In the second panel (frame B), the objects are in the relative lower positions. The insets in the Step 2 show the extracted slices from the upper and lower positions, distinguishing between A and B spectra. These 2D spectra are then displayed in the Step 3 part of the diagram, after the rectification. Step 4 shows the A-B and B-A results for both upper and lower slices (the colour bars here are inverted for a better visual inspection of the H$\alpha$ knots). Finally, Step 5 shows the final 2D spectrum obtained from the combination of the Step 4 intermediate products.  }
\label{datareduction}
\end{figure*}

\section{Data reduction}\label{datareductionsect}
Near-infrared multi-object spectroscopic data have been reduced with a custom set of routines written by the lead author and optimised to reduce LUCI observations with curved-slit MOS configuration. These routines have been extensively tested, comparing the final 2D spectra of the observed lensed galaxies with the ones reduced with an independent pipeline, the ``flame'' routine (\citealt{Belli2017}), created specifically for the LUCIs instruments. 

Here the main steps of the data reduction are briefly summarised (see also Fig. \ref{datareduction}).
After the removal of cosmic rays using the L.A. Cosmic procedure (\citealt{vanDokkum2001}), we corrected for the non-linear response of the detector following the prescriptions described in the LUCI User Manual\footnote{Available for download at \url{http://abell.as.arizona.edu/~lbtsci/Instruments/LUCIFER/lucifer.html}}. We separated the frames belonging to the A and B positions, respectively, using a dedicated slit with a star devoted to tracing the position of each frame on the detector y-axis (the spatial direction; the spectral direction is along the horizontal axis; see Fig. \ref{datareduction}, Step 1). We measured the reference star position using the spatial profile constructed integrating the stellar continuum within a 100-pixel sky-line-free interval along the x-axis. From a Gaussian fit of the spatial profile, we then derived the star position (from the Gaussian centre) and the spatial resolution (from the FWHM) associated with the individual frame. 
We then extracted from the MOS image the A and B two-dimensional (2D) spectra slices within the edges of the curved slits, taking into account the dithering pattern (Fig. \ref{datareduction}, Step 2). 

We corrected the curvature of the 2D spectra by performing a rectification so that the spatial and spectral channels were all aligned in dispersion and spatial pixel space (Fig. \ref{datareduction}, third panel). This step is based on the identification of individual bright sky emission lines, which can be reliably used to trace the variations in the spectral dispersion pixel scale along the Y axis. 
Specifically, for a given frame, we extracted a single-row spectrum $F_{\rm Y^*}(x)$ from the scientific 2D spectrum at Y =Y$^*$ as a reference for the spectral dispersion pixel scale; we then selected a set of prominent sky lines and derived their central positions via Gaussian fitting, constructing the vector $C_{\rm Y^*}=\{ x_i\}$, with $x_i$ being the pixel position of the centroid of the i-th sky line. We repeated the process for a second single-row spectrum extracted at the position Y$^*+$j along the Y-axis, considering the same set of prominent sky lines and obtaining the vector $C_{\rm Y^*+j}$. The variations in the dispersion axis with respect to the position Y$^*$ are corrected performing a polynomial fit of order 2 of the centroid pixel position vectors $C_{\rm Y^*}$ and $C_{\rm Y^*+j}$. Finally, we used the obtained polynomial function to interpolate the $F_{\rm Y*+j}$(x) spectrum. We repeated this process for different j values, covering the extension of the curved 2D spectrum.

Due to grating instability during the acquisition of subsequent frames, we referred all the rectified spectra for a given instrument configuration to the spectral dispersion pixel scale of the first scientific frame. 

We removed the sky lines and background emission by subtracting the 2D frames of the B dither position from the A dither position (and vice versa), pairing subsequent 
 frames (Fig. \ref{datareduction}, Step 4). We coadded all the A sky-subtracted spectra and all the B sky-subtracted spectra separately. Finally, after a wavelength calibration performed using OH sky lines, we combined the A and B spectra. 

Flux calibrations were performed using the 2MASS magnitudes of the reference stars. For J1038+4849 (J1038 hereinafter) and J0022 we used emission line fluxes derived from the integral field unit (IFU) spectroscopic data published by Jones et al. (2012; for J1038) and the long-slit spectra by Finkelstein et al. (2009; for J0022) to test the reliability of our calibration: very good agreement  (within $\lesssim 20\%$; see Table \ref{spectralfitresults}) was found comparing integrated emission line fluxes from the same images covered by their observations.
 We note that slit losses cannot be excluded in our sample: some emission falls outside the slits because of seeing effects and the inaccuracy of the perfect centring on the curved slits along the entire extension of the lensed sources. Nevertheless, for J0022 and J1038, the comparison between ARGOS and previous observations shows that they are not significant. 

 In
Table \ref{tablog}, we report the median spatial resolutions derived from our observations. We note that lower spatial resolutions were usually measured in individual frames, with FWHM values as small as 0.25-0.30 arcsec. In this work, however, we decided to use the entire set of frames to gain higher-S/N spectra; as a result, the final angular resolution is slightly larger. 
  
The final 2D spectra we obtained from the data reduction are reported in Figs. \ref{integratedspectra} and \ref{integratedspectra2}. 
In the following sections we briefly discuss the general spectroscopic analysis. 

\begin{figure*}[t]
\centering
\includegraphics[width=8.2cm,trim=0 10 0 0,clip]{{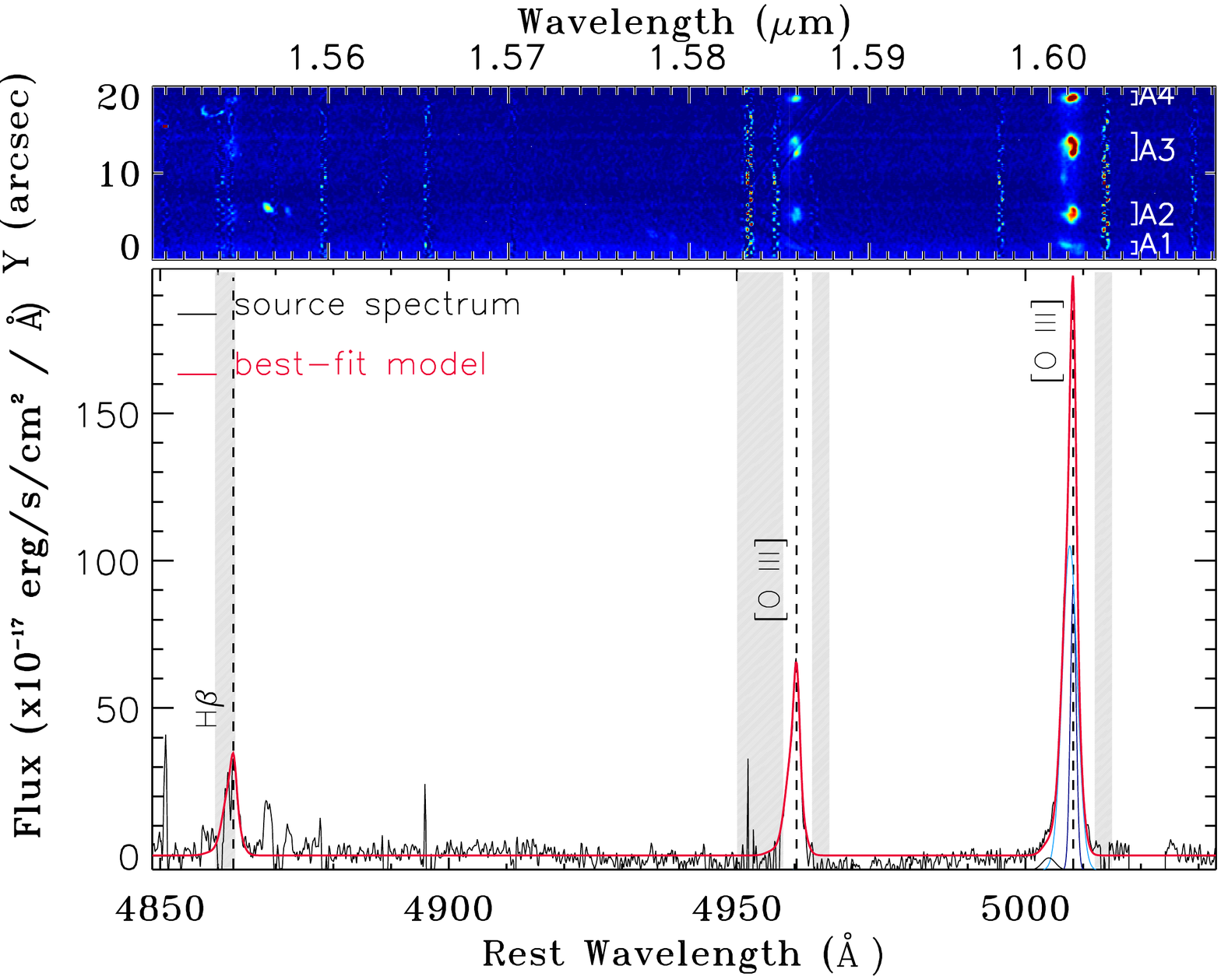}}
\hspace{0.01cm}
\includegraphics[width=8.2cm,trim=0 10 0 0,clip]{{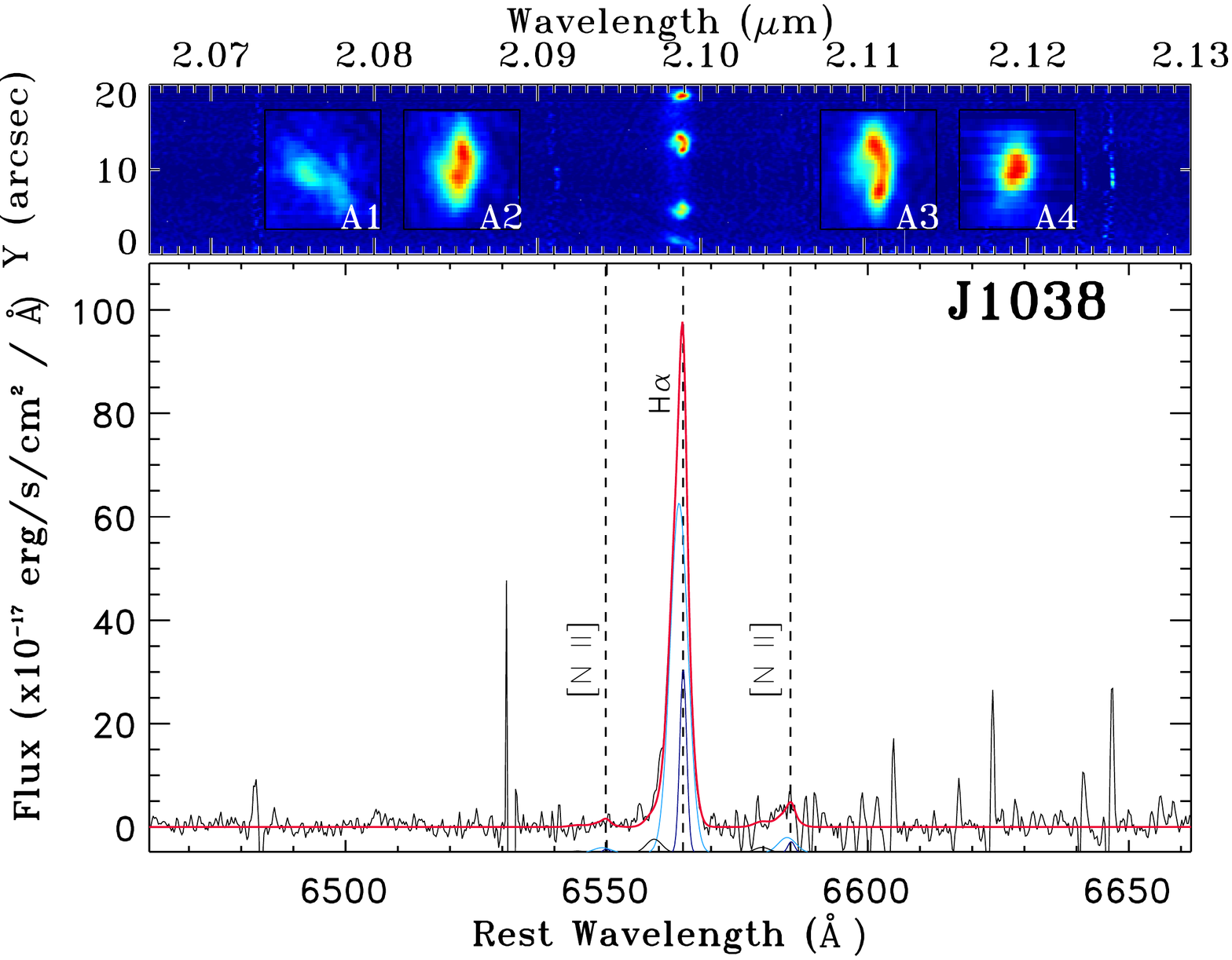}}
\hspace{0.01cm}
\includegraphics[width=8.2cm,trim=0 10 0 0,clip]{{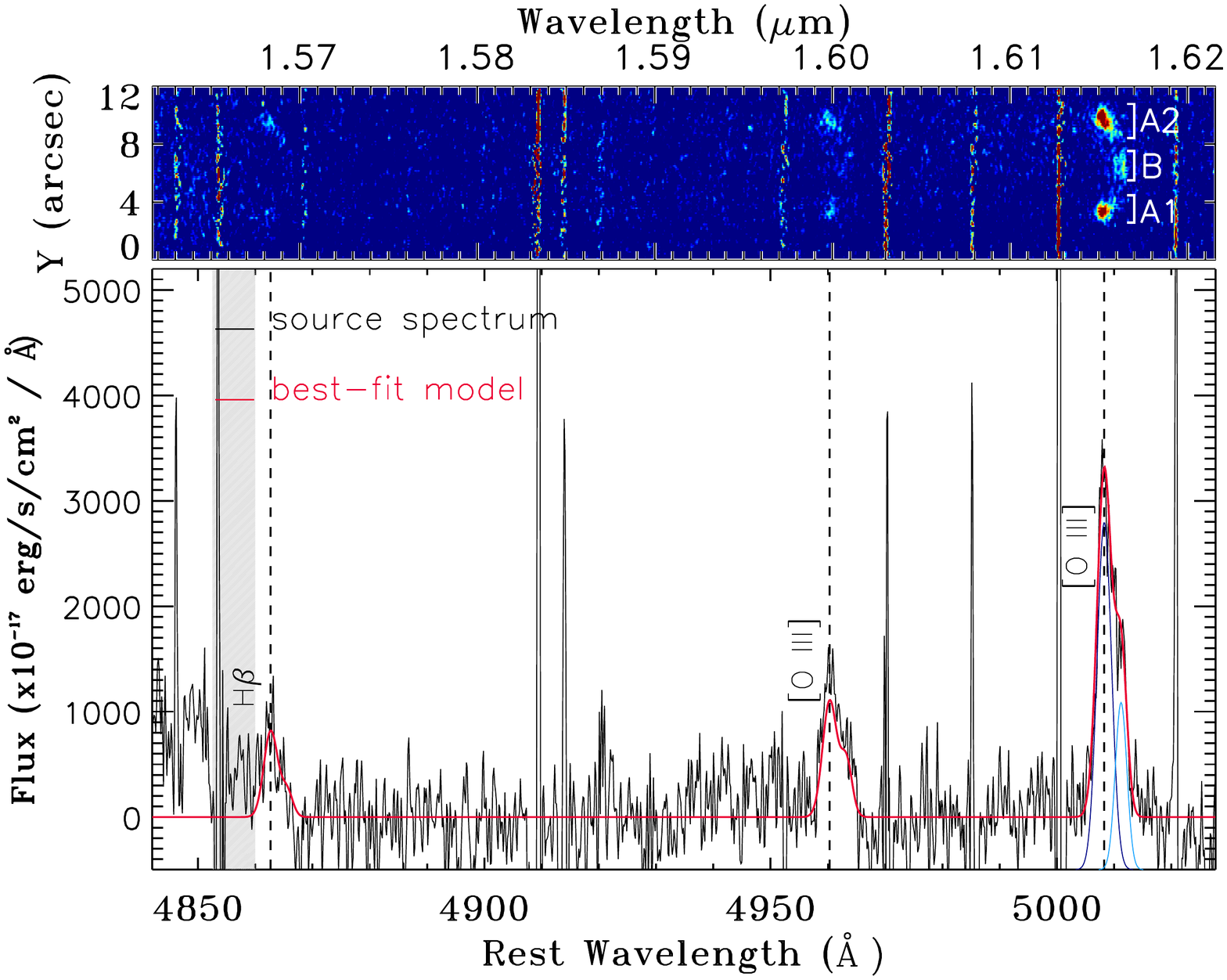}}
\hspace{0.01cm}
\includegraphics[width=8.2cm,trim=0 10 0 0,clip]{{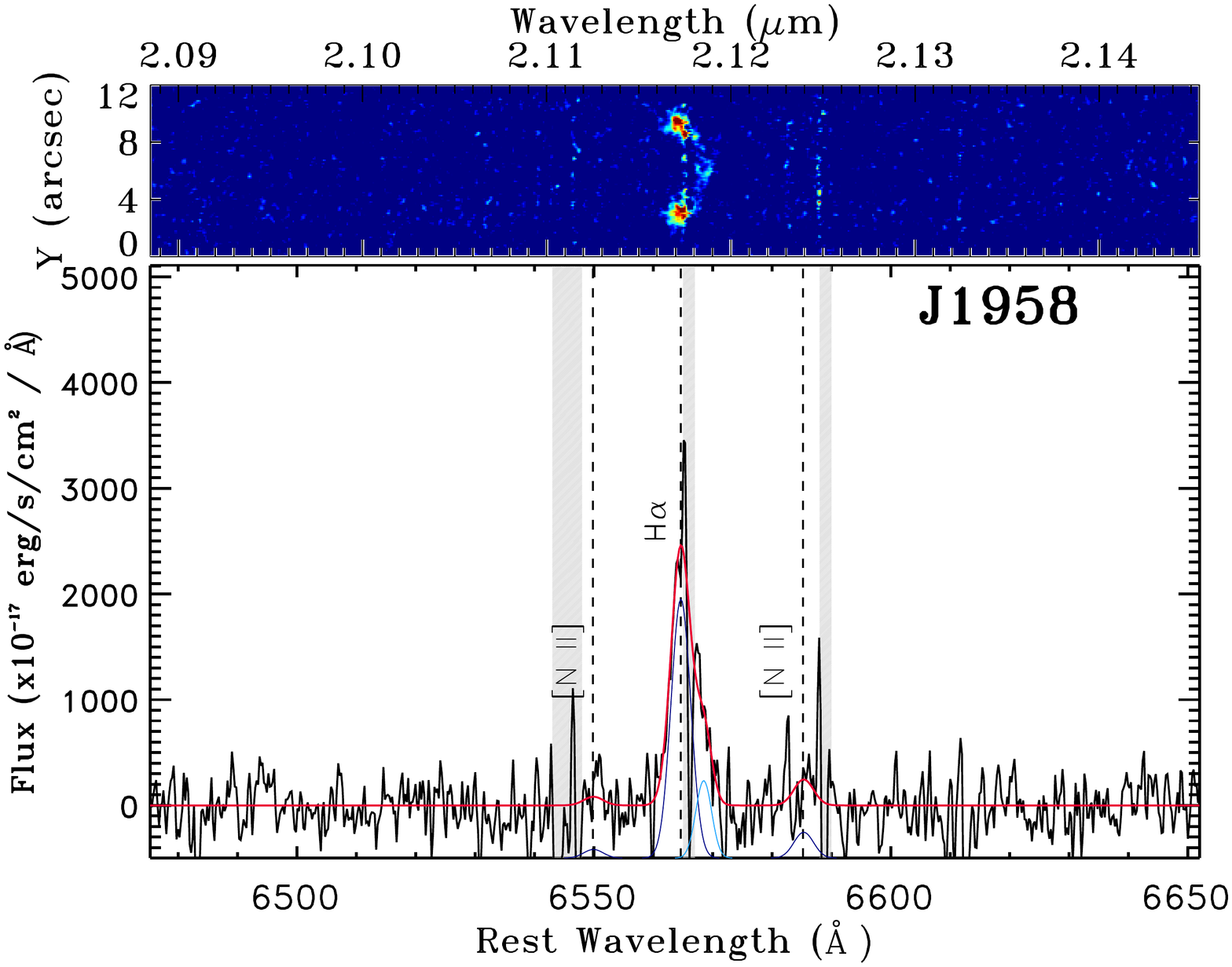}}
\hspace{0.01cm}
\includegraphics[width=8.2cm,trim=0 10 0 0,clip]{{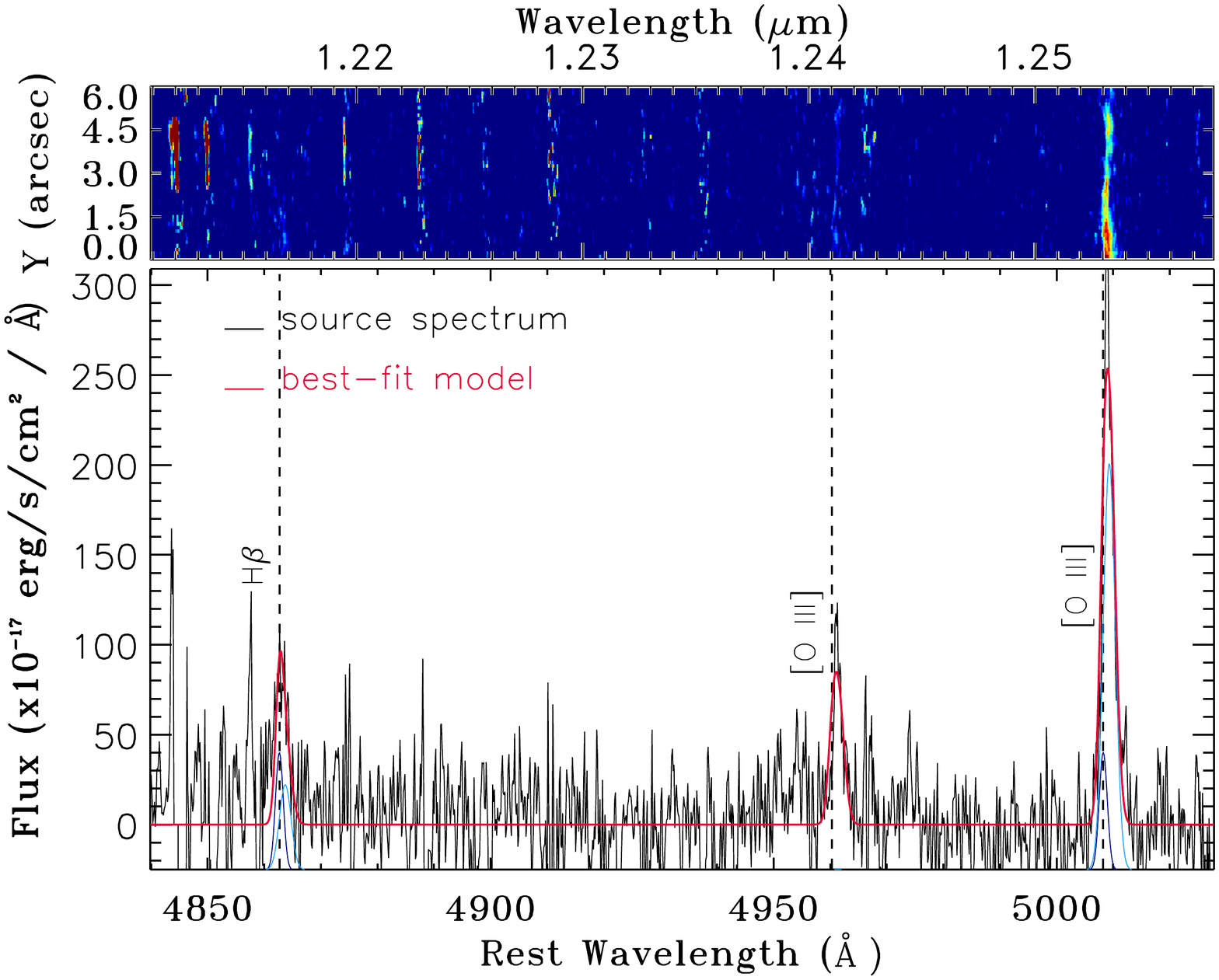}}
\hspace{0.01cm}
\includegraphics[width=8.2cm,trim=0 10 0 0,clip]{{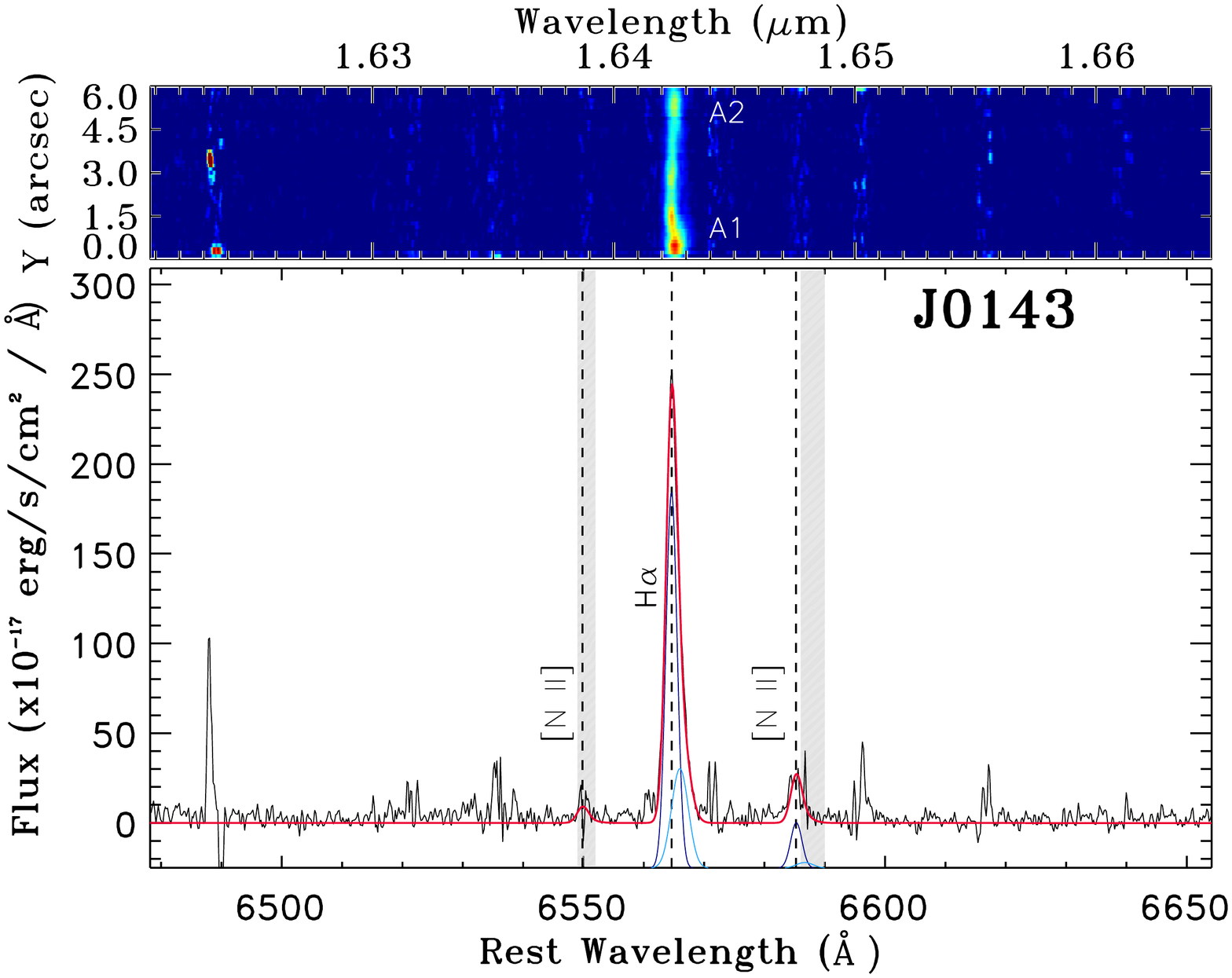}}
\caption{\small 
 Portion of LBT/ARGOS NIR integrated spectra of J1038 (top panels), J1958 (central panels) and J0143 (bottom panels). Superimposed on the spectra are the best-fitting components presented in Section \ref{fit}, with arbitrary normalisation for visualisation purposes; the red solid curve shows the sum of all best-fit components. The dotted lines mark the wavelengths of H$\beta$, [O III]$\lambda\lambda$4959,5007 (left) and [N II]$\lambda$6548, H$\alpha$ and [N II]$\lambda$6584 (right), from left to right, respectively.  The regions excluded from the fit of the Gaussian components and corresponding to the most intense sky lines are highlighted as shaded areas. For each spectrum and each region, the upper panel shows the observer-frame 2D spectrum (blue to red colours represent increasing fluxes).  The closing square bracket labels in the 2D spectra locate the regions (spaxels) from which we obtained the 1D integrated spectra associated with the different knots. For J1038, we also show zoomed portions of the 2D spectrum associated with the four different H$\alpha$ knots. The features at $\sim 1.558 \mu$m in the 2D spectrum of J1038 are associated with the LUCI detectors' persistences.
}

\label{integratedspectra}
\end{figure*}

\begin{figure*}[ht]
\centering

\includegraphics[width=8.2cm,trim=0 4 0 2,clip]{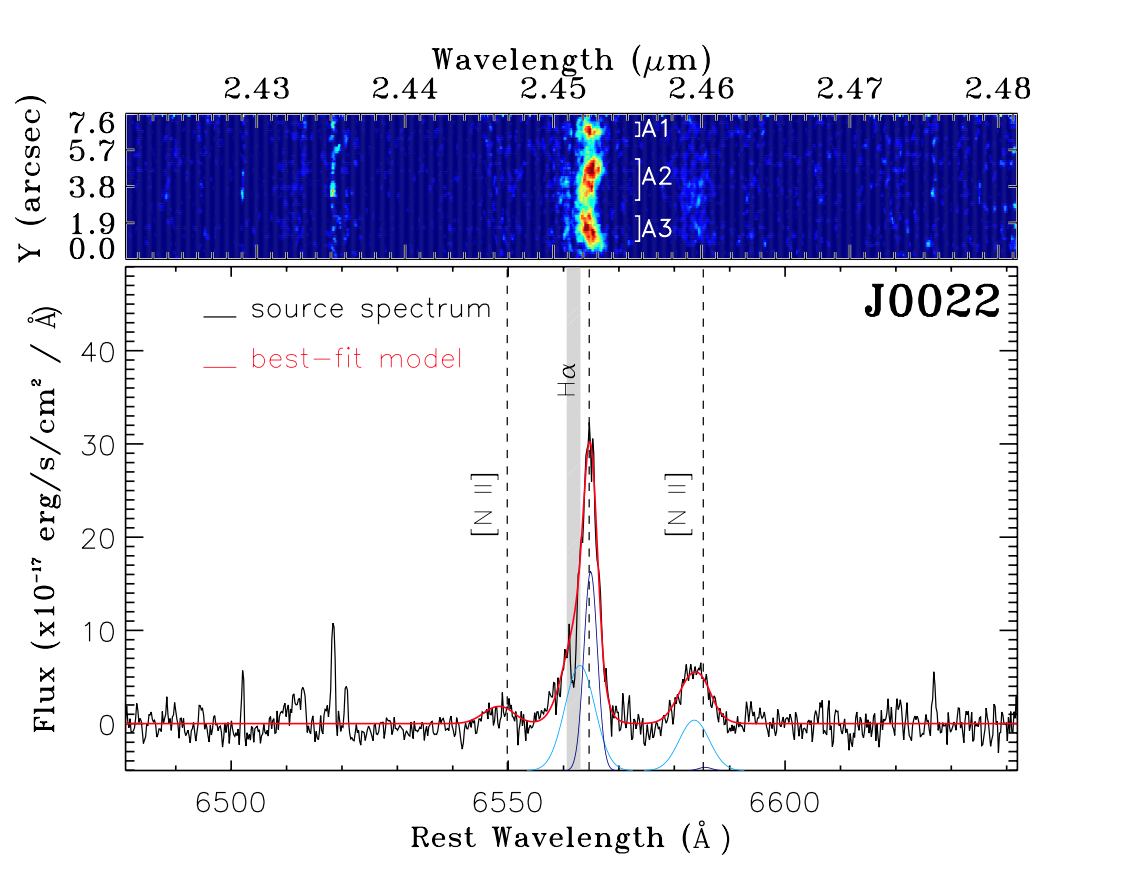}
\includegraphics[width=8.2cm,trim=5 8 0 0,clip]{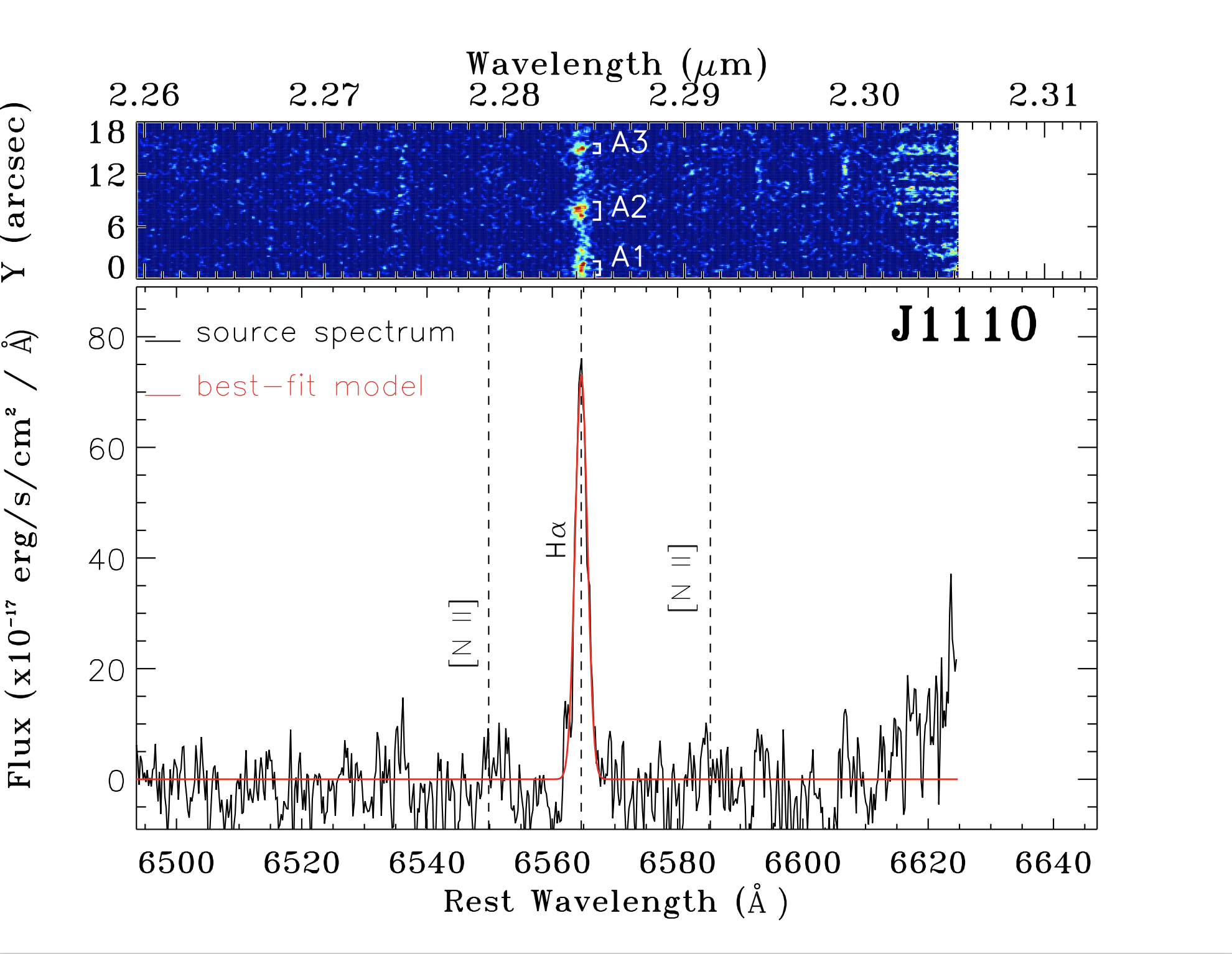}
\includegraphics[width=8.2cm,trim=5 0 0 5,clip]{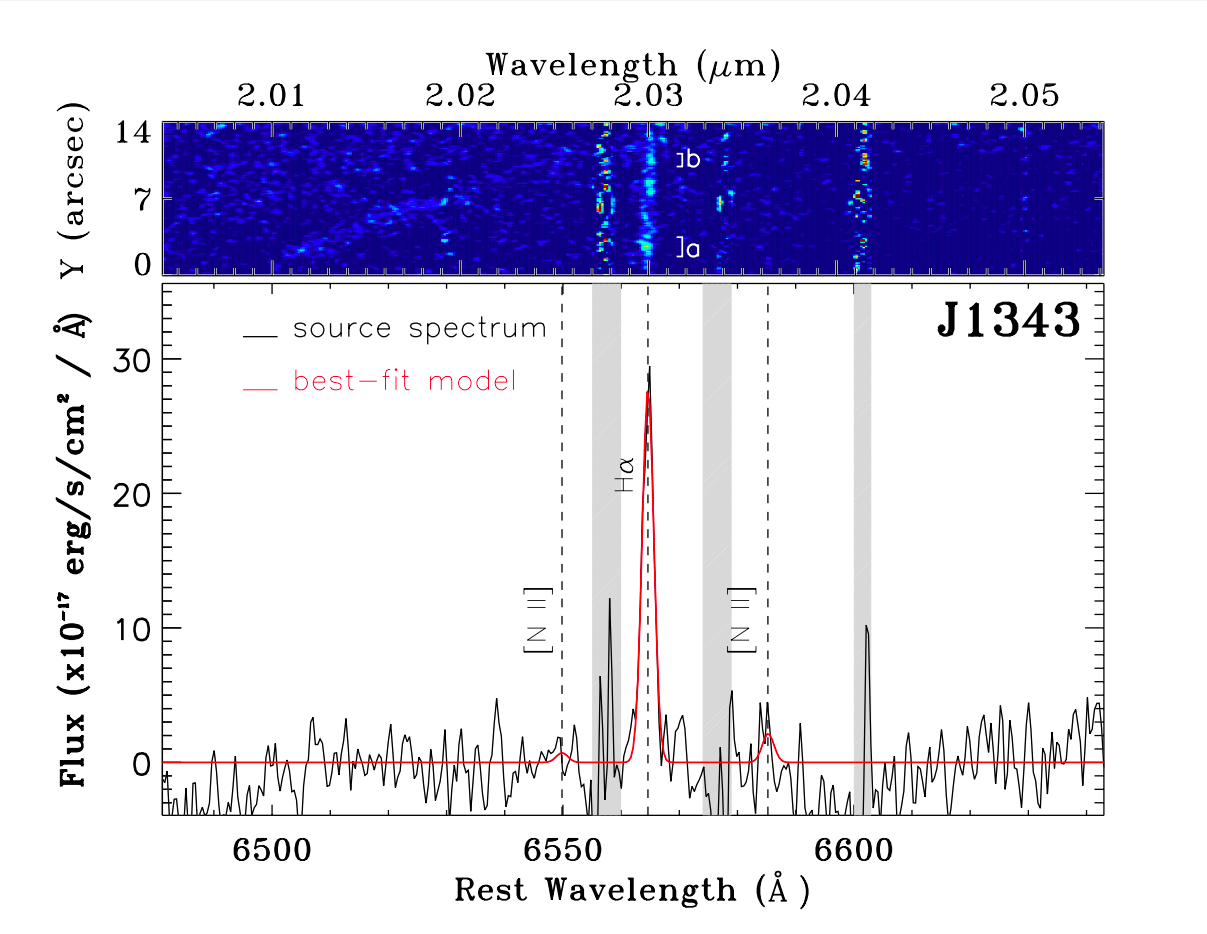}
\caption{\small Portion of LBT/ARGOS K-band integrated spectra of J0022 (top left), J1110 (top right) and J1343 (bottom). See Fig. \ref{integratedspectra} for detailed description.
}
\label{integratedspectra2}
\end{figure*}

\section{Spectroscopic analysis}\label{fit}


Figures \ref{integratedspectra} and \ref{integratedspectra2} show the one-dimensional (1D) integrated rest-frame spectra, extracted from $6-18''$ apertures, according to the presence of distinct blobs and extended arc-like emission line systems along the spatial direction in the 2D spectra. 
The aperture tracing along the dispersion axis, for each lensed galaxy,  was performed using the relative `reference star' as a reference and applying the same tracing to the target spectra.

For each source, we defined the systemic redshift on the basis of the best fitting solution which gives the wavelengths of narrow components of detected rest-frame emission lines in the integrated spectra. Spectral fits were performed following the prescriptions described in our previous studies (e.g. \citealt{Brusa2015}). 
We used LevenbergMarkwardt least-squares fitting code MPFITFUN (\citealt{Markwardt2009}) to reproduce all the detected optical emission lines with Gaussian profiles. 
Due to the presence of asymmetric line profiles, different sets of Gaussian profiles were used to reproduce the J1958, J0022, J0143 and J1038  emission lines.

 For each set of Gaussians, we imposed the following constraints: the wavelength separation between emission lines is constrained in accordance to atomic physics; the flux ratios between [N {\small II}]$\lambda$6583 and  [N {\small II}]$\lambda$6548 and between  [O {\small III}]$\lambda$5007 and  [O {\small III}]$\lambda$4959 components is fixed to 2.99:1 (\citealt{Osterbrock2006}). We excluded from the fit the regions where significant sky-subtraction residuals downgrade the emission line profiles (e.g. Fig. \ref{integratedspectra}, grey shaded area).

To best characterise the kinematic properties in all our targets in a homogeneous way, we used the non-parametric approach (e.g. Rupke \& Veilleux 2013). In fact, this approach does not depend on the number of Gaussian components used to model the emission lines, which 
may suffer from strong degeneracy between the different components (see e.g. \citealt{Zakamska2014}). 

Non-parametric velocities are obtained measuring the velocity $v$ at which a given fraction of the total best-fit line flux is collected using a cumulative function. The zero point of velocity space is defined adopting the systemic redshift derived from the integrated spectra. We carried out the following velocity measurements on the best fit of [O {\small III}]$\lambda$5007 and/or H$\alpha$ line profiles, depending on the wavelength coverage of LBT observations and the absence of bad sky-subtraction residuals affecting the line profiles:

\begin{itemize}
\item
$W80$: The line width comprising 80\% of the flux, defined as the difference between the velocity at 90\% ($V90$) and 10\% ($V10$) of the cumulative flux;
\item
$V10$: The maximum blueshift velocity parameters, defined as the velocity at 10\% of the cumulative flux; 
\item
$V90$: The maximum redshift velocity parameters, defined as the velocity at 90\% of the cumulative flux; 
\item
$Vpeak$: The velocity associated with the emission line peak;
\item
$V50$: The velocity associated with 50\% of the cumulative flux. 
\end{itemize}

These velocities are usually introduced to characterise asymmetric line profiles. In particular, $W80$ is closely related to the velocity dispersion for a Gaussian velocity profile ($W80\sim 1.1\times$ FWHM), and is therefore indicative of the line widths, while  maximum velocities ($V10$ and $V90$) are mostly used to constrain outflow kinematics when prominent line wings are found (e.g. \citealt{Cresci2015, CanoDiaz2012,Perna2017a}). $Vpeak$ has been used as a kinematic tracer for rotating disk emission in the host galaxy (e.g. \citealt{Harrison2014, Brusa2016}): the peak flux density of the emission line can trace the brightest narrow emission-line components, when perturbed gas emission contribution is not dominant.

Table \ref{spectralfitresults} summarises the relevant spectral fit measurements; reported errors were estimated from Monte Carlo trials of mock spectra (see, e.g. \citealt{Perna2017a} for details). 
In the following sections, we briefly introduce the properties of the lensed galaxies and describe the detailed spectral analysis for each target.

\subsection{J1038, the Cheshire cat arc}\label{j1038}

\begin{figure*}[t]
\centering
\includegraphics[width=9.cm,trim=60 70 60 20,clip]{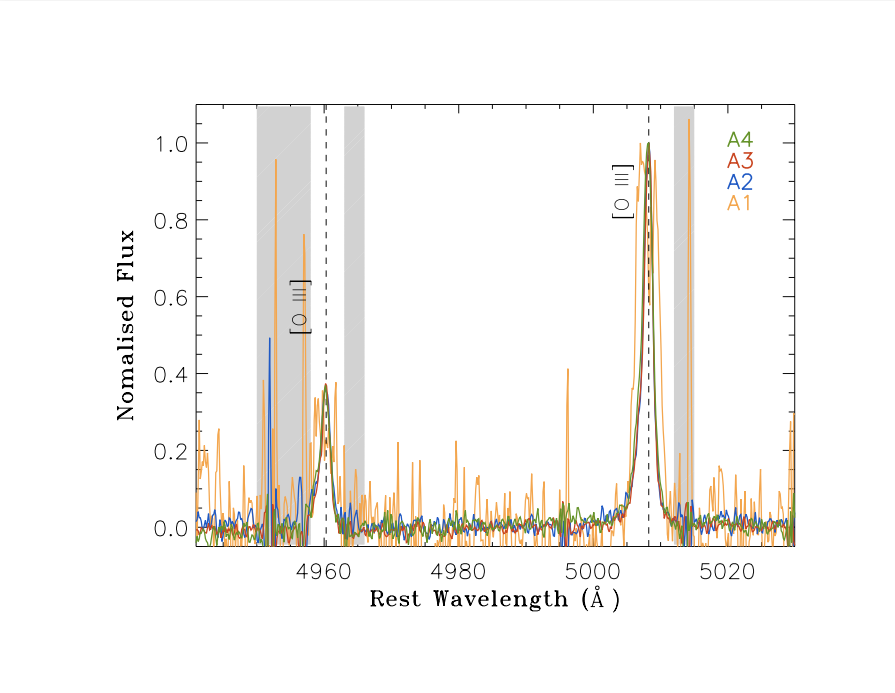}
\hspace{0.01cm}
\includegraphics[width=9.cm,trim=60 70 60 20,clip]{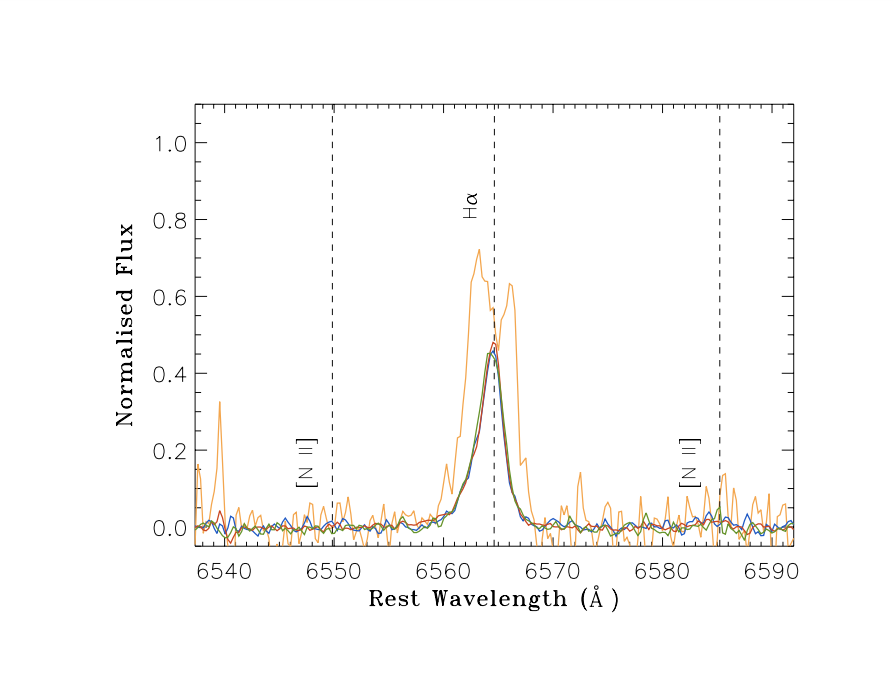}
\caption{\small J1038 integrated spectra of the four bright knots, colour-coded as labelled in the left panel. Grey shaded regions mark the wavelength ranges most affected by sky-line residuals and excluded from the spectral fit. To facilitate the comparison, we normalised the fluxes so that the peak intensity of the [O III] line is unity for each integrated spectrum.
}
\label{integratedspectraJ1038}
\end{figure*}

\begin{figure}[h]
\centering
\includegraphics[width=9.5cm,trim=90 380 16 60,clip]{{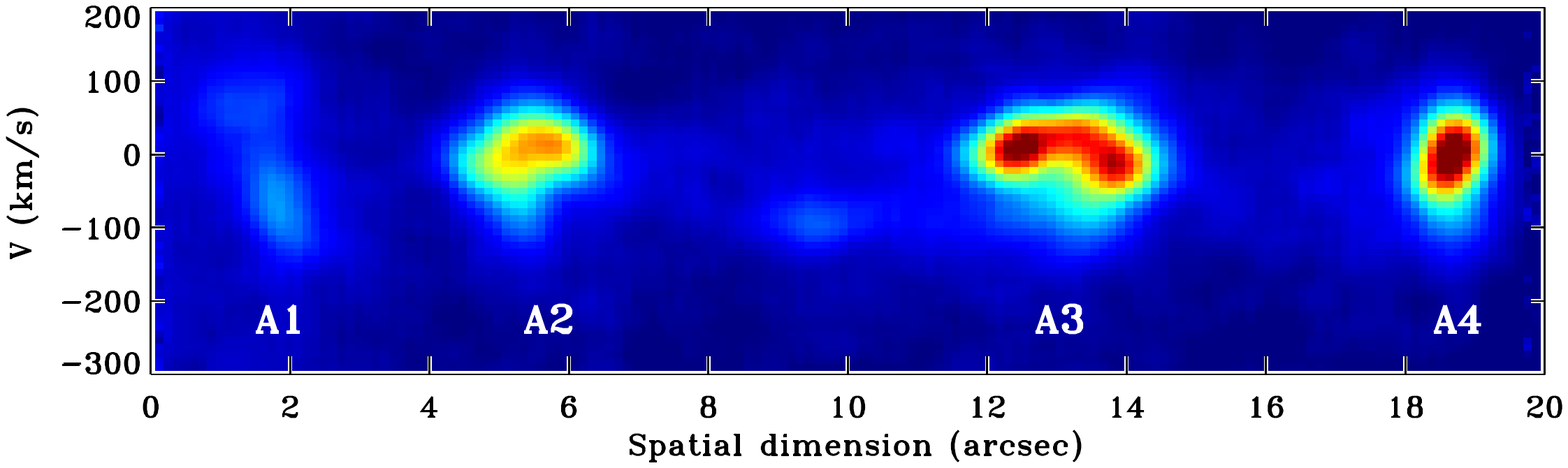}}
\includegraphics[width=9.5cm,trim=10 30 10 20,clip]{{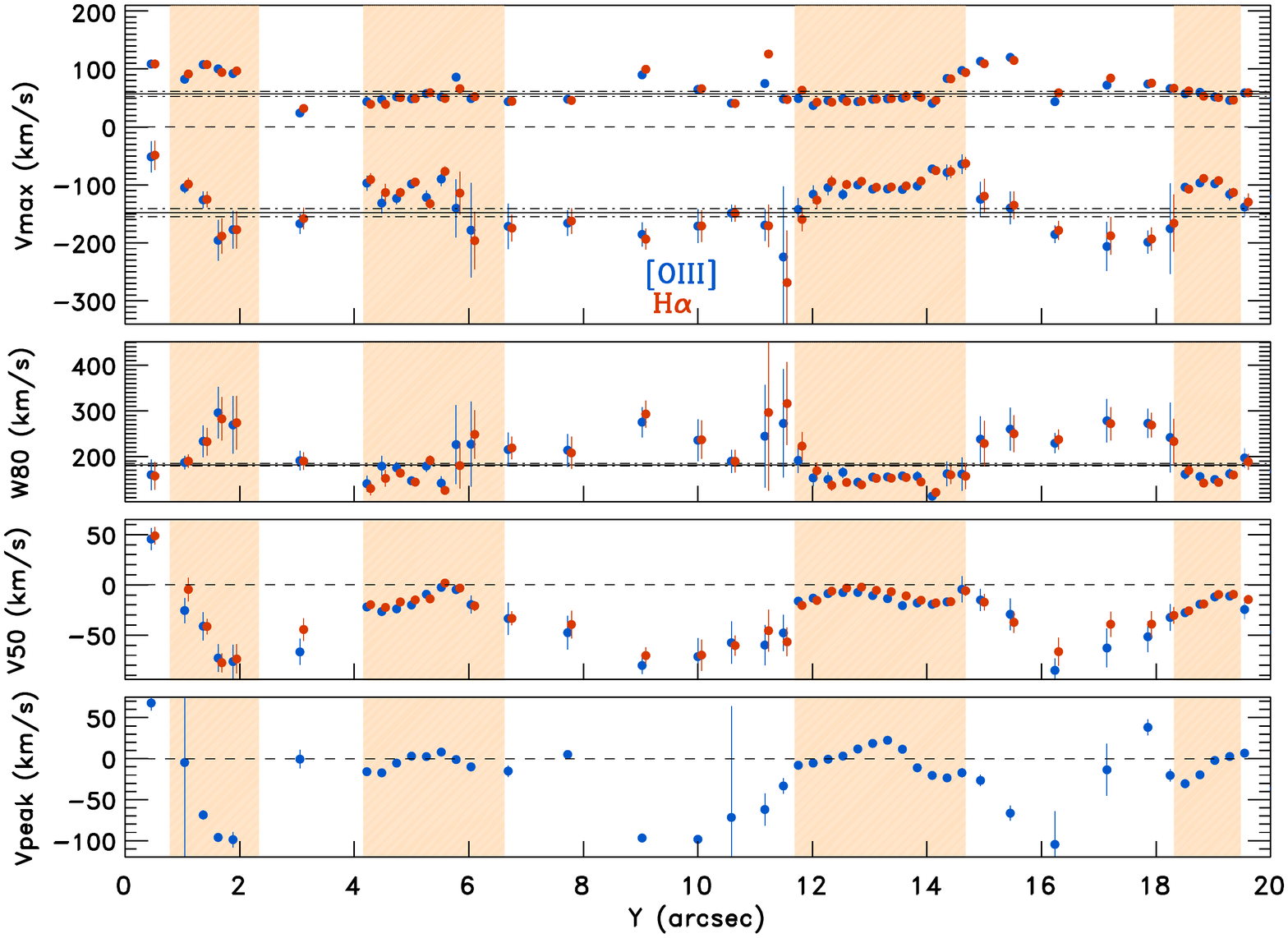}}
\caption{\small {\it Top panel:}  zoom of the  J1038 2D spectrum around the [O {\small III}] line. {\it Lower panels:} non-parametric velocity measurements as a function of the position along the spatial axis from which 1D spectra have been extracted. Red and blue symbols refer to H$\alpha$ and [O III]$\lambda$5007 line velocity measurements. Orange shaded areas refer to the regions associated with the knots in 2D spectra, as labelled in the top panel. In the first panel, V90 (positive) and V10 (negative) measurements are reported. The other panels show $W80$, $V50,$ and $Vpeak$ variations.  For $V10$, $V90$, and $W80$, we show with solid lines the average velocities derived from the total integrated spectrum for the [O III] line (dot-dashed lines refer to $\pm 1\sigma$). For $V50$ and $Vpeak$ panels, we indicate the 0 km/s velocity with dashed lines. We refer to  Sect.  \ref{fit} for the description of individual non-parametric velocity estimators (and associated distinct kinematic properties).
}
\label{nonparametricJ1038}
\end{figure}

SDSS J1038+4849 is a lensed system associated with a relatively poor group of galaxies at z $\sim 0.45$, dominated by two luminous elliptical galaxies. \cite{Belokurov2009} and \cite{Bayliss2011} identified four major ring systems associated with background galaxies at redshift 0.80, 0.97, 2.20 and 2.78 (see also \citealt{Irwin2015}). This system is also known as the Cheshire cat (\citealt{Carroll1866}) gravitational lens because of the smiling-cat-like appearance defined by the spatial configuration of the foreground and background lensed galaxies.  

Archival observations taken with HST WFC3 using the F390W, F110W, and F160W filters are available for this target. 
Part of the extended arc structure has been observed with the NIR integral field spectrograph OSIRIS at Keck II (see Fig. \ref{images}), and K-band spectroscopic results have been presented by \cite{Jones2013}. 

ARGOS observations cover  the entire arc associated with the galaxy at z $=2.20$ and previously observed with Keck II. Figure \ref{integratedspectra} (top panels) shows the integrated spectrum around the H$\beta$+[O {\small III}] and H$\alpha$+[N {\small II}] lines, redshifted in H- and K-band, respectively. We fitted all the emission lines simultaneously using atomic physics constraints and the same kinematic components for each set of Gaussian profiles. Due to the presence of strong sky line residuals affecting the blue wing of the H$\beta$, we constrained the amplitude ratio of the different Gaussian sets to be the same for the two Balmer lines. The presence of prominent extended blue wings is noticeable in Balmer lines and [O {\small III}] profiles, with associated velocities as high as $V10= -148\pm 7$ km/s (to be compared with the smaller $V90 = +57\pm 3$ km/s). These profiles are generally interpreted as due to outflows. In fact, while maximum velocities as high as $\sim$ 1000 km/s are easily observed in local and high-redshift galaxies associated with intense AGN activity (e.g. \citealt{Bischetti2017,Genzel2014,Kakkad2016,Lanzuisi2015,Mullaney2013,Zakamska2016}), much smaller velocities are expected for SF-driven outflows (e.g. \citealt{Cicone2016, Cresci2017, Lagos2013, Murray2005,Newman2012,Talia2017}). In Sect. \ref{fluxratios}, we report classical flux ratio diagnostics (\citealt{Baldwin1981}) to study excitation conditions of the emitting gas, for both unperturbed and unsettled components, in order to decipher whether or not the physical conditions are compatible with outflows.

HST images show that this arc-like structure is composed of multiple overlapping arcs and knots.  Three luminous knots are easily distinguishable (A2, A3 and A4 in Fig. \ref{images}), and a fainter image (A1) can be identified at the lower end of the arc. 

A source reconstruction through lens modelling has been presented by \citet{Jones2013}. According to these authors, the A1 and A2 images do not correspond to the same emitting region in the source plane. Moreover, A2, A3, and A4 are all associated with the same source-plane region, with the parity of A3 inverted with respect to those of A2 and A4 (Jones, private communication).

All the knots detected in HST images are seen in ARGOS spectra, especially in [O {\small III}] and H$\alpha$ lines (see Fig. \ref{integratedspectra}, top panels). A3 is the more elongated structure and presents two small sub-systems both in HST images and NIR 2D spectra, with the north one showing greater extension at shorter wavelengths. The same sub-systems may be tentatively recognised also in A2 and A4 images (see the insets in the top right panel), but with blueish extension in the south sub-components. These kinematically resolved shapes in [O {\small III}] and H$\alpha$ line knots in 2D spectra confirm that A2, A3, and A4 are different images of the same physical region, with A3 inverted with respect to the others,  in line with the \citet{Jones2013} lens model. 
The shape of A1 is instead quite different. 
Finally, we also note the presence of a small faint knot between the A2 and A3 images, noticeable in the [O {\small III}]$\lambda$5007 emission line and fairly resolved in H$\alpha$, blueshifted with respect to the bulk of the doubly ionised emission.

In Fig. \ref{integratedspectraJ1038}, we report the spectra extracted from the regions in the proximity of the brightest knots associated with [O {\small III}]$\lambda$5007 and H$\alpha$ (we consider the regions above $5\sigma$ as contour levels; see labels in 2D spectra, Fig. \ref{integratedspectra}). We also extracted the spectrum from the fainter blob (A1) considering the regions above $2\sigma$ as contour levels. 
The spectral analysis revealed that A2, A3, and A4 are characterised by consistent (within $1\div 2\sigma$) kinematic properties, with [O {\small III}] $W80$ velocities of the order of $155\pm 4$ km/s (see Table \ref{spectralfitresults} for more details).
The spectrum associated with A1, although affected by lower S/N, is instead significantly different: broad ($W80([O {\small III}]) = 202\pm 5$ km/s), double peaked profiles are recognised in the more prominent emission lines.  Moreover, we notice that the values of $f_{[O III]}/f_{H\alpha}$ and $f_{[O III]}/f_{H\beta}$ are consistent, given the observational uncertainties, for A2, A3, and A4, while A1 shows significantly different flux ratios, as expected. 

We also note that the [O {\small III}] and H$\alpha$ fluxes of the source A3 are consistent with the sum of the fluxes associated with A2 and A4 ($f_{A2}\approx f_{A3}+f_{A4}$; see Table \ref{spectralfitresults}). According to the cusp relation between the magnifications of multiple images (e.g. \citealt{Keeton2005}), when the source lies close to a cusp caustic, the three closely spaced images of the source should satisfy $\left | \mu_{a} \right | - \left | \mu_{b} \right | +\left | \mu_{c} \right |=0$, where $\mu_b$ is the signed magnification of the central image. This relation can explain the observed fluxes, in line with the fact that A3 is inverted with respect to A2 and A4.  Therefore, our results agree with the lens model presented by \citet{Jones2013}, according to which the images A2, A3, and A4 are associated with a unique emitting region in the source plane while A1 refers to different sources (Jones, private communication).

In order to study the sub-systems in the bright lensed galaxy images,  we derived kinematic maps of the [O {\small III}]$\lambda$5007 and H$\alpha$ emission lines analysing 1D spectra extracted along the Y axis. A minimum S/N of 5 was required for the detection of the emission lines; if this criterion was not met for a single spaxel, the adjacent spaxels were averaged until both  [O {\small III}]$\lambda$5007 and H$\alpha$ lines were detected with S/N $>5$. The lower panels in Fig. \ref{nonparametricJ1038} show the non-parametric velocity measurements for individual 1D spectra; in the upper panel, we reported a zoom of the 2D spectrum around the [O {\small III}] line, to highlight the complex kinematics associated with the rest-frame optical emission lines.

The region associated with A1 presents peculiar velocity gradients, with increasing $W80$ and $|V10|$  going from south to north (low to high Y values in Fig. \ref{nonparametricJ1038}); the velocity peak, instead, goes from positive to negative values, similar to the $V50$ measurements. All these trends are due to the presence of double peaked profiles (see  Fig. \ref{integratedspectraJ1038}, orange profiles), with an increasing flux contribution of the bluest component going from lower to higher Y values (see also Fig. \ref{nonparametricJ1038}, top panel). The other bright knots are instead associated with mild velocity variations, with $V10 \approx -100$ km/s, $V90 \approx 50$ km/s, and $W80 \approx 150$ km/s. The presence of asymmetric profiles with prominent blue wings in both  [O {\small III}] and H$\alpha$ lines is highlighted by the difference of a factor of two between $V10$ and $V90$ measurements. Regular variations are instead observed in the panels showing $V50$ and $Vpeak$ measurements, going from negative to positive and then again to negative values with increasing Y in A2, A3, and A4. The fact that the more A3 blueshifted regions appear at higher Y, in contrast to what happens in A2 and A4, again confirms the fact that A3 is inverted with respect to the other images.

Non-parametric measurements also show the presence of highly disturbed kinematics in the overlapping arcs between the knots, with $V10 < -150$ km/s and $W80 > 200$ km/s, possibly associated with shocks/outflows in the galaxy.  These velocities are larger than the ones in proximity of the bright knots. This may be explained assuming that the outflowing gas is more extended than the bulk of emission, associated with systemic, unperturbed material (see the source reconstruction presented by \citealt{Jones2013}): under this assumption, in fact, the regions between the knots the non-parametric velocities are less affected by the systemic emission component. Similar behaviour has also been observed in the ionised component of a multi-phase massive outflow of an obscured QSO at z $\sim 1.5$, XID 2028 (\citealt{Brusa2017,Cresci2015,Perna2015a}). The outflow characterisation of J1038 is discussed in Sect. \ref{outflowmassrateeta}.

\cite{Jones2013} observed the southern component of the J1038 arc, covering the A2 image and the upper part of A1. They observed a large velocity offset between these two images and concluded that they are the components of a merging system with a mass ratio ($6 \pm 3$):1 (deriving the stellar masses from spectral energy distribution (SED) analysis), with the A1 component being the more massive and luminous in IR bands.
The source plane reconstruction presented by \citealt{Jones2013} shows the presence of faint H$\alpha$ emission in regions between the two companions. This emission could be associated with ionised gas in tidal streams as well as with outflowing gas (see Fig. 3 of \citealt{Jones2013}). In fact, the presence of a companion at $\sim 6$ kpc does not exclude the presence of outflows in the system (see e.g. \citealt{Rupke2013,Rupke2015, Feruglio2013, Saito2017, Harrison2012, Vayner2017, Banerji2017} for examples of interacting systems in which evidence of outflows has been found, both in the local Universe and at z$\sim 2$).

The high spatial resolution provided by ARGOS and the use of curved-slit matching the entire extension of the arc allowed us to observe the entire kinematic structure associated with A1. 
We note that, while the kinematic signatures found in A1 are not observed in other regions across the arc, disturbed kinematics are observed over the entire extension of the lensed system. We therefore suggest that the presence of galaxy-wide outflows may be a more plausible scenario to explain the observed kinematics in all images apart from A1. Moreover, the presence of  substructures in the A2, A3, and A4 images may suggest that the lensed galaxy has a clumpy morphology, as usually observed in high-z galaxies, and the presence of outflows may be associated with these star forming clumps (e.g. \citealt{Genzel2014}).

Summarising, we find evidence of disturbed kinematics possibly associated with outflows over the entire extension of the arc, with the exclusion of the region associated with the A1 image, related to a different emitting region (e.g. a galaxy companion; \citealt{Jones2013}) dominated by rotating motions.

\subsection{J1958}\label{j1958}

\begin{figure}[t]
\centering
\includegraphics[width=9.5cm,trim=90 380 16 60,clip]{{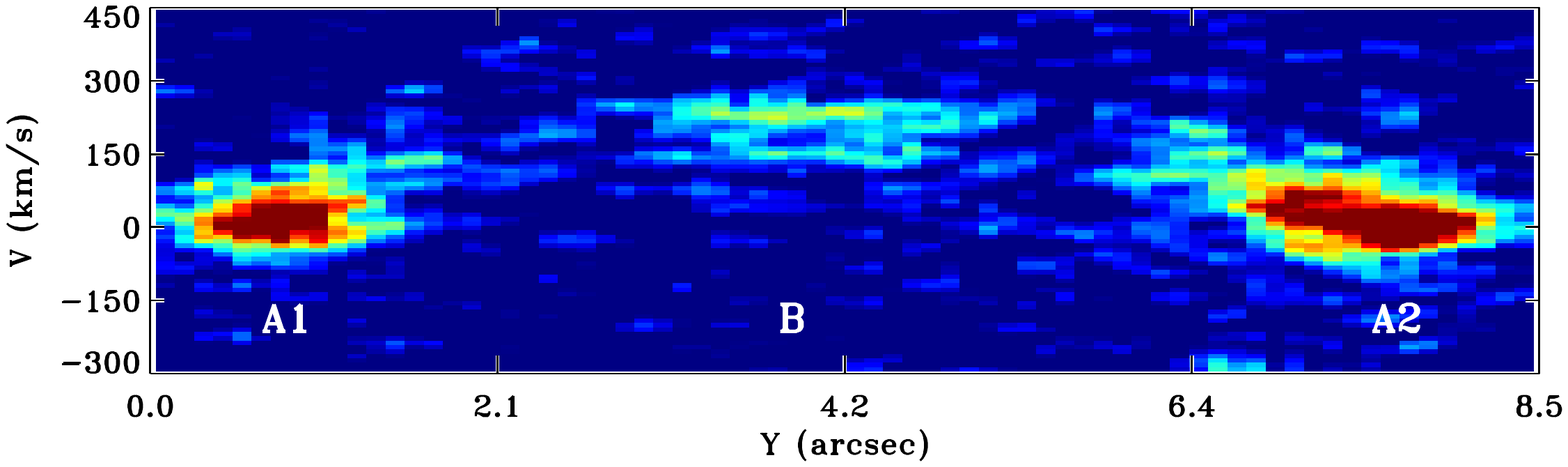}}
\includegraphics[width=9.5cm,trim=10 10 10 20,clip]{{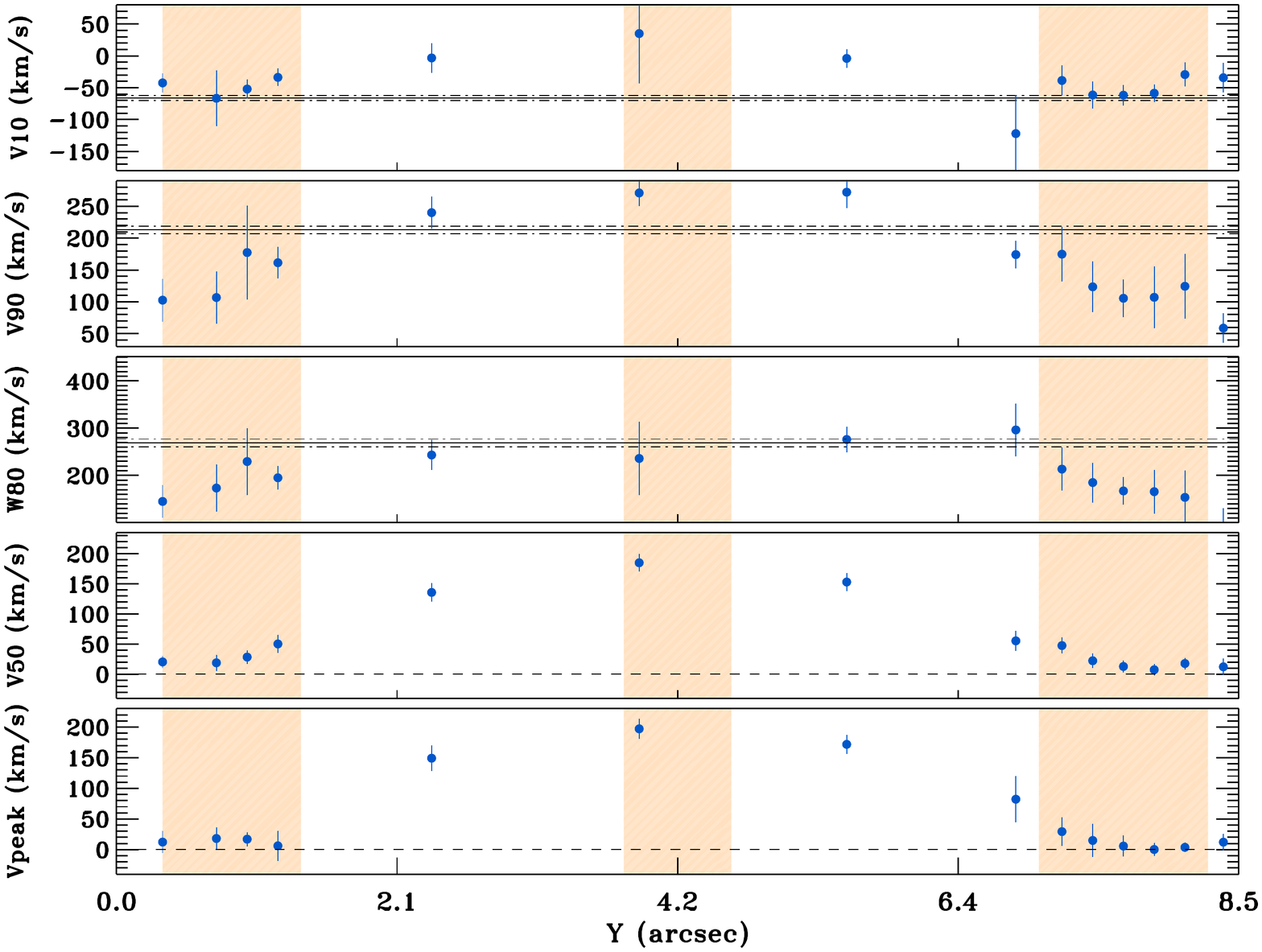}}
\caption{\small {\it Top panel:}  zoom of the J1958 2D spectrum around the [O III] line. {\it Lower panels:} J1958 non-parametric velocity measurements as a function of the position along the dispersion axis from which 1D spectra have been extracted. Blue symbols refer to [O III] measurements. See Fig. \ref{nonparametricJ1038} for a more detailed description of the figure.
}
\label{nonparametricJ1958}
\end{figure}

SDSS J1958+5950, the giant arc at redshift z $=2.225$, is lensed by a rich galaxy cluster at redshift 0.24 and was discovered by \cite{Stark2013}. No HST images are available for this source; we obtained an ARGOS image in Ks-band on the commissioning night of May 19, 2016, using LUCI 1, with an integration time of 74 minutes (Fig. \ref{images}). The system is comprised of three images and extends over $\gtrsim 5'' $. 

Figure \ref{integratedspectra} (central panels) shows the integrated spectra around the H$\beta$+[O {\small III}] and H$\alpha$+[N {\small II}] lines, redshifted in H- and K-band, respectively. Because of the low S/N and the presence of intense sky lines in the proximity of Balmer lines, we constrained,  also for this target, the relative amplitudes of Gaussian profiles reproducing the H$\beta$ and H$\alpha$ lines. 
The best-fit shows the presence of two distinct Gaussian components of each emission line. These components are also easily visible in the 2D spectra (insets in the central panels of Fig. \ref{integratedspectra}). 

A lens model for the J1958 system has been presented by \citet{Stark2013}. It was recently used by \cite{Leethochawalit2016}: the authors, using IFU OSIRIS observations covering the three images of J1958 (CSWA128 in their work), observed that the two kinematic components, also detected in the ARGOS spectra, are actually spatially separated  (by $\sim 1-2$ kpc) in the source plane. Moreover, they found that the region between the two kinematic components is associated with the largest velocity dispersions (see Fig. 1 in \citealt{Leethochawalit2016}). According to these results, the target can be associated with a merger system.

As for the target previously presented, we compared the spectra of A1
and A2 images and found evidence that they are related to the same emitting region in the source plane ( non-parametric velocities are consistent within $1\div 2\sigma$; Table \ref{spectralfitresults}). The central image (B), barely distinguishable in the Ks-band LBT image (Fig. \ref{images}), appears instead redshifted by $191\pm 13$ km/s (with $\gg 4\sigma$ significance).

Due to the presence of bad sky-subtraction residuals around Balmer lines, we derived spatially resolved kinematic analysis only for the doubly ionised oxygen. In Fig. \ref{nonparametricJ1958}, we reported the kinematic map of the [O {\small III}]$\lambda$5007 emission line derived from the analysis of 1D spectra extracted along the Y axis. The variations in all non-parametric velocities display a symmetric behaviour, with increasing velocities going toward the centre of the slit. The peak-to-peak velocity amplitude  between the A1/A2 images and the central B knot is $\sim 200$ km/s, while maximum velocities vary by a factor of two; taken together, it is unlikely that these changes can be associated with internal motions within the gravitational potential of a galaxy. Alternatively, this may be suggestive of the presence of two different lensed galaxies caught during an ongoing interaction, as also reported in \citet{Leethochawalit2016}. Finally, we note that both velocity dispersion and peak velocity variations reported by  \citet{Leethochawalit2016} are consistent with those derived in this work.

\subsection{J0143}\label{j0143}

The system SDSS J0143+1607 was discovered as  part of the CASSOWARY survey (\citealt{Belokurov2009}). It is lensed by a source at z $= 0.42$ and comprises a bright arc and one counter image associated with a source at redshift 1.502. The spectroscopic information of the two sources was derived by \cite{Stark2013}, analysing long-slit optical spectra obtained at the Multiple Mirror Telescope. The spatial resolution of the
SDSS {\it griz} images, the only available imaging data for this system, does not allow us to infer the geometry of the lensed galaxy within the bright arc; for visual purposes, however, we indicated with A1 and A2 the emission coming from the opposite ends of the curved slit, possibly associated with two distinct knots (Fig. \ref{images}).  The spectroscopic ARGOS observations cover the entire bright arc with a curved slit of  $0.7''$ in width.

Simple galaxy-galaxy lens models have been reported for this target by \citet{Stark2013} and \citet[][CSWA 116 in their papers]{Kostrzewa2014}. The source plane reconstruction shows an extended source close to the cusp, with the nuclear regions outside of the caustic. This configuration is compatible with the presence of a bright arc in the image plane, associated with a double image of the nuclear regions (Kostrzewa-Rutkowska, private communication).

In the bottom panels of Fig. \ref{integratedspectra}, we report the integrated spectrum of J0143 around the H$\beta$+[O {\small III}] and H$\alpha$+[N {\small II}] lines, redshifted in the J- and H-bands, respectively. The best-fit model reported in the figure was obtained with the simultaneous multi-component approach. Asymmetric lines with extended red wings were modelled with two sets of Gaussian profiles. The presence of bad sky-line residuals strongly affects the [N II]$\lambda$6549 line and the blue wing of the [N II]$\lambda$6584 feature. These regions have therefore been excluded from the fit. We also note that in the J-band spectrum, the target is off-centre by $\sim 0.5''$ toward the lower part of the curved slit (see the 2D spectra in Fig. \ref{integratedspectra}); therefore, that part of the A1 region was not perfectly sampled. 

The 2D spectrum shows a clumpy morphology in the line emission, with at least four different knots associated with different line centroids along the dispersion axis (see also Fig. \ref{nonparametricJ0143}, top panel). The bottom panels of Fig. \ref{nonparametricJ0143}  report the kinematic map of the H$\alpha$ line for this target, analysing 1D spectra extracted along the Y axis. Disturbed kinematics are observed in the regions A1 and A2, with $V90\sim 100$ km/s (to be compared with corresponding $V10 \sim 50$ km/s;  see Fig. \ref{nonparametricJ0143} for typical errors). These measurements may be interpreted as due to the presence of receding high-velocity gas driven by stellar winds.

\begin{figure}[h]
\centering
\includegraphics[width=9.5cm,trim=90 380 16 60,clip]{{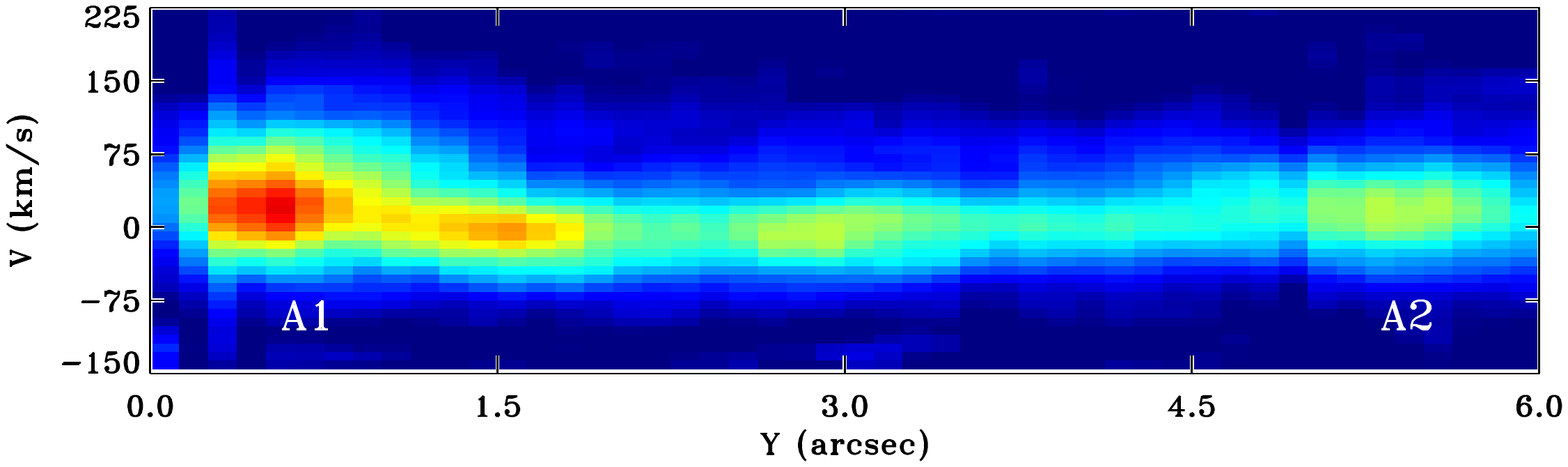}}
\includegraphics[width=9.5cm,trim=10 30 10 20,clip]{{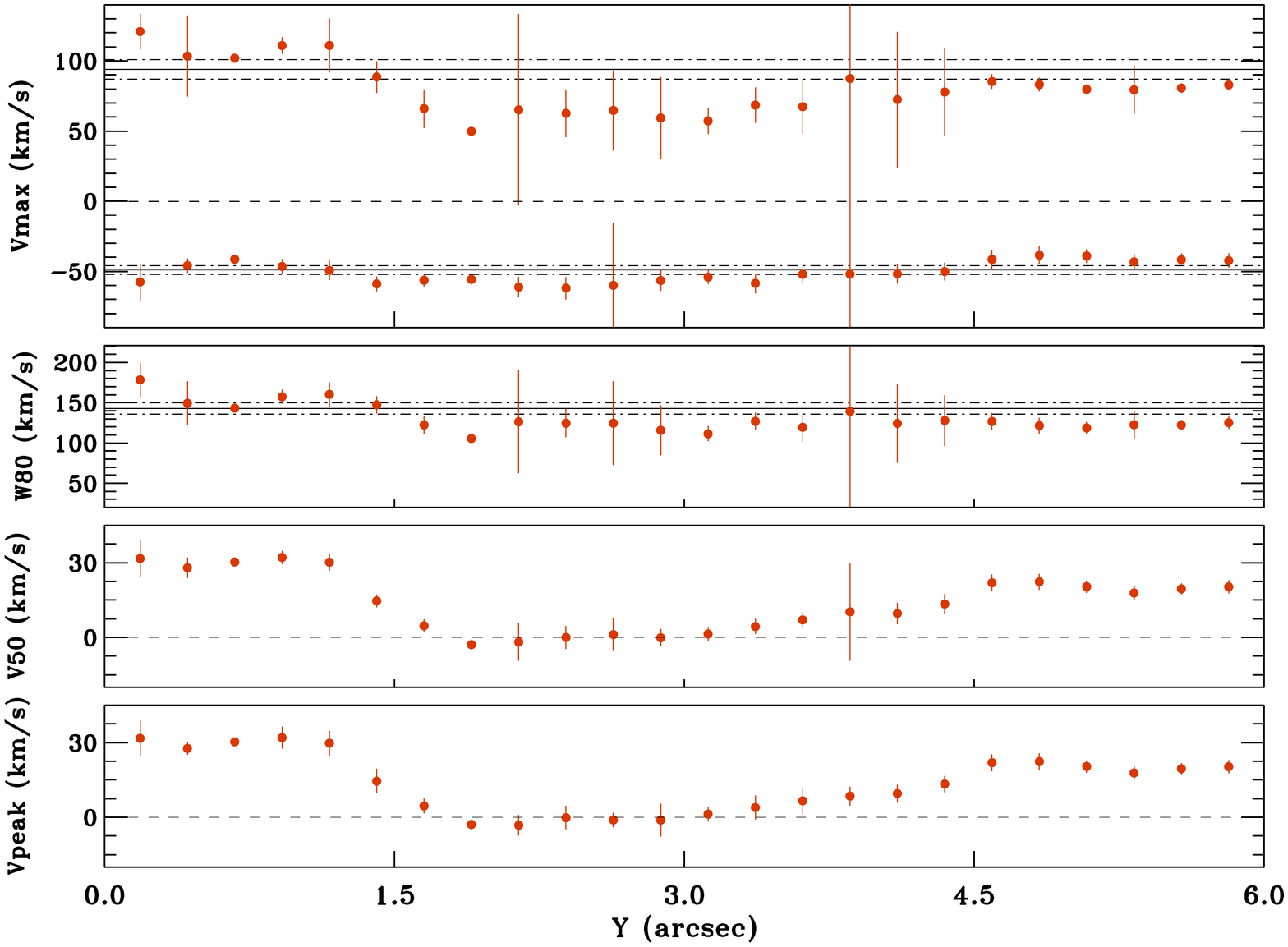}}
\caption{\small {\it Top panel:}  zoom of the J0143 2D spectrum around the H$\alpha$ line. {\it Lower panels:} J0143 non-parametric velocity measurements as a function of the position along the spatial axis from which 1D spectra have been extracted. Red symbols refer to H$\alpha$ line velocity measurements. In the first panel, V90 (positive) and V10 (negative) measurements are reported. The other panels show $W80$, $V50$, and $Vpeak$ variations.  See Fig. \ref{nonparametricJ1038} for a more detailed description of the figure.
}
\label{nonparametricJ0143}
\end{figure}

\subsection{J0022, the 8 o'clock arc}\label{j0022}

\begin{figure}[h]
\centering
\includegraphics[width=9.5cm,trim=90 380 16 60,clip]{{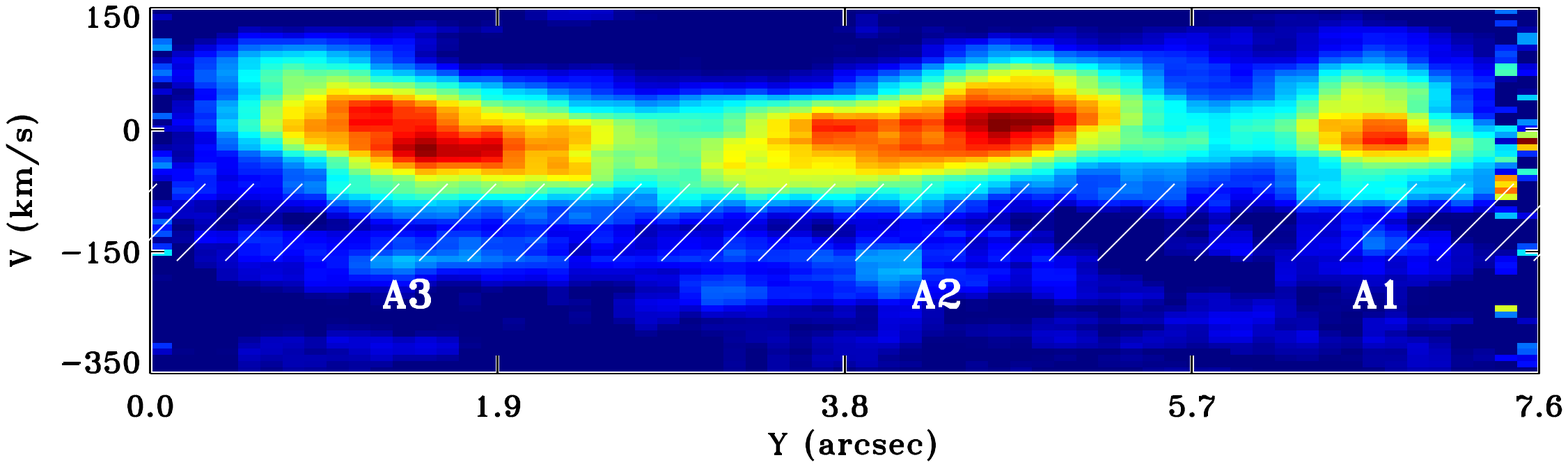}}
\includegraphics[width=9.5cm,trim=10 30 10 20,clip]{{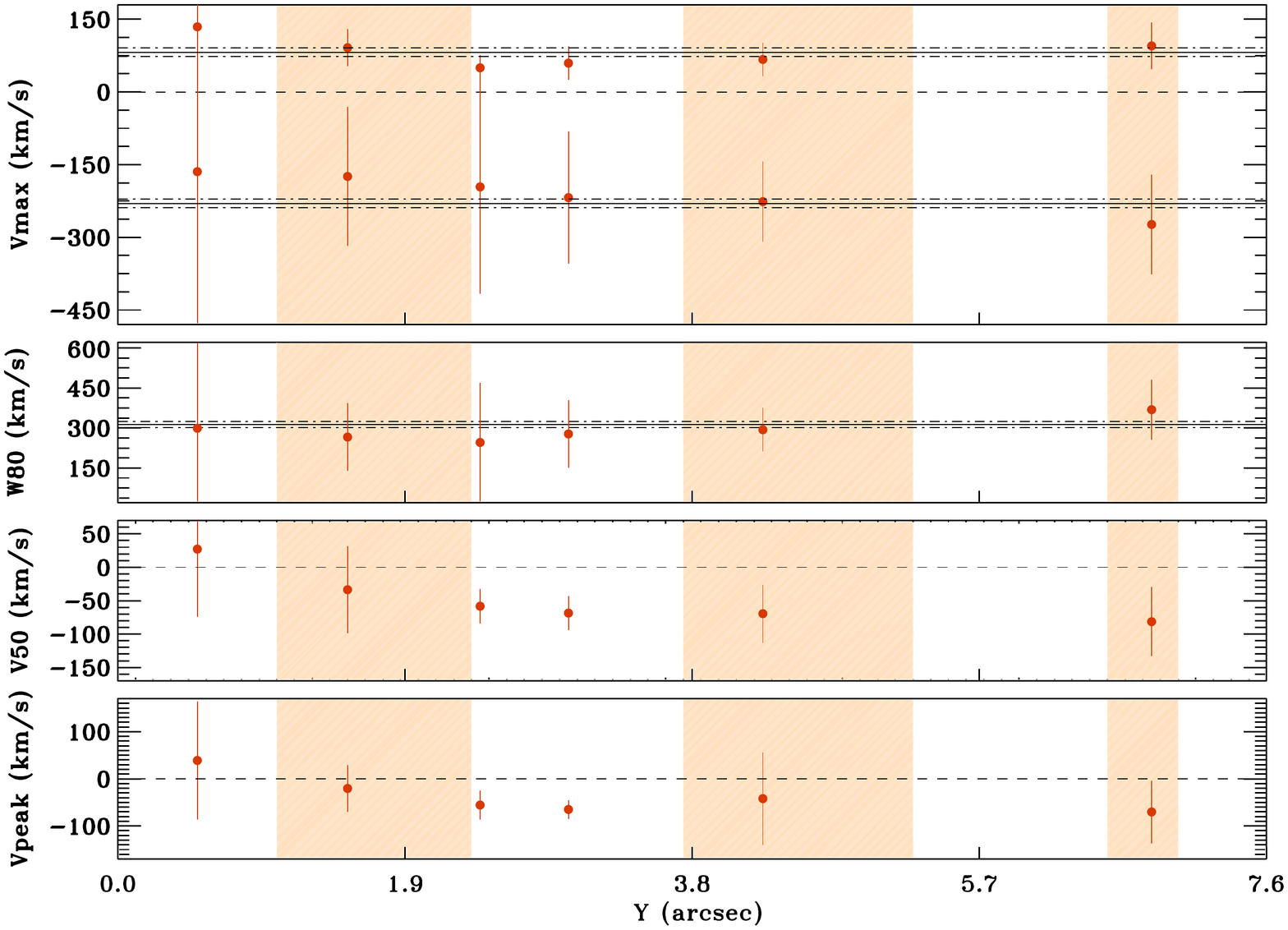}}
\caption{\small {\it Top panel:}  zoom of the J0022 2D spectrum around the H$\alpha$ line; the shaded white area shows the region associated with bad sky-line subtraction residuals. {\it Lower panels:} J0022 non-parametric velocity measurements as a function of the position along the spatial axis from which 1D spectra have been extracted. Red symbols refer to [N II] line velocity measurements. In the first panel, V90 (positive) and V10 (negative) measurements are reported. The other panels show $W80$, $V50,$ and $Vpeak$ variations.  See Fig. \ref{nonparametricJ1038} for a more detailed description of the figure.
}
\label{nonparametricJ0022}
\end{figure}

J0022+1431 lensed galaxy was discovered by \cite{Allam2007}, from visual inspection of an SDSS imaging data catalogue of candidate interacting/merging galaxy pairs. Follow-up imaging and spectroscopy revealed that this system, dubbed by Allam et al. the "8 o'clock arc", is associated with a bright Lyman break galaxy at z $=2.73$ strongly lensed by a z $=0.38$ red galaxy (\citealt{Allam2007}).
The system has been observed with HST/NICMOS (using the F110W and F160W filters; see Fig. \ref{images}), and with HST/WFC2 (F450W, F606W, F814W).  Long-slit spectroscopic follow-up observations have also been obtained covering part of the elongated structure with Keck LRIS (optical) and Gemini NIRI (NIR; \citealt{Finkelstein2009}), and with VLT/X-Shooter (UV to NIR; \citealt{Dessauges2010,Dessauges2011}). Integral field NIR spectroscopic data have been obtained at VLT/SINFONI, mapping for the first time the entire extension of the arc in H-band (\citealt{Shirazi2014}). The study of the most prominent optical emission lines of J0022 has however been limited by observational constraints: the [O {\small III}]$\lambda$5007 line falls outside the H band ($\lambda_{obs} \sim 1.87\mu$m), while the H$\alpha$ is redshifted at $\sim 2.45 \mu$m in the vicinity of a strong atmospheric sky-line (see \citealt{Dessauges2011}). 

The top-left panel of Fig. \ref{integratedspectra2} shows the LBT/ARGOS integrated spectrum over the entire curved-slit, with our best-fit model superimposed. The presence of bad sky-line residuals strongly affects the blue wing of the H$\alpha$ line. We therefore used the [N {\small II}]$\lambda$6584 line to derive non-parametric estimates for this target, obtaining $W80 = 313 \pm 12$ km/s and $V10 = -230 \pm 9 $ km/s.

In the top panel of Fig. \ref{nonparametricJ0022},  we show a zoom of the LBT/ARGOS 2D spectrum of J0022 in the proximity of the H$\alpha$ line. The Balmer emission is associated with two prominent knots in the lower and central parts of the slit (Y $\sim 1.5''$ and $\sim 4''$ in the figure, respectively), which are connected by an extended arc; an additional knot is also distinguishable in the upper part of the slit (Y $\sim 7''$). Overall, the elongated structure and the three images can be related to the arc shown in  the HST image (Fig. \ref{images}). 
The H$\beta$ flux map of the arc obtained by \cite{Shirazi2014} using the IFU instrument SINFONI shows relative fluxes between the three images which are consistent with those in the LBT/ARGOS spectrum. Therefore, we can confirm that differential slit losses are not at the origin of the different fluxes associated with the three knots.

Gravitational lens models of the arc have been presented by \citet{Dessauges2011} and  \citet{Shirazi2014}; the three images are related to a unique emitting region in the source plane, with A2 inverted with respect to the other two images. For consistency with the analysis of J1038 and J1958, for this
galaxy we also compared the integrated spectra extracted from the three different knots, deriving, as expected, consistent non-parametric measurements (within 1$\sigma$), with $W80\approx 280$ km/s and $V10 \approx -200$  km/s (see Table \ref{spectralfitresults}).

The presence of distinct kinematic components is notable from the 2D
spectrum in Fig. \ref{nonparametricJ0022}, showing  faint Balmer emission at high negative velocities ($\lesssim -150$ km/s) over the entire extension of the lensed galaxy. 
The bad sky-line residuals affecting the blue wing of the H$\alpha$ line, however, do not allow for the use of the bright Balmer line for the study of kinematic variations along the Y axis. We therefore used the fainter [N {\small II}]$\lambda$6584 line to perform spatially resolved kinematic analysis. In Fig. \ref{nonparametricJ0022} (lower panels), we report the non-parametric velocity measurements obtained from the analysis of 1D spectra extracted along the Y axis, following the same prescriptions as described in the previous sections (but requiring for each 1D spectrum a line detection at S/N $>3$). All non-parametric velocity measurements are consistent within 1$\sigma$ along the arc. 
We note however that, despite the large uncertainties associated with these measurements, the [N {\small II}] non-parametric velocities reflect the variations in the Balmer line profile in the 2D spectrum: the bulk of the emission coming from the overlap of the A1 and A2 images is shifted to more negative velocities and is responsible for the small variation observed in $V90$. Instead, the emission associated with faster-approaching gas seems to be uniformly distributed over the extension of the lensed galaxy ( $V10= -230\pm 9$ km/s from the total integrated spectrum). Possible interpretations of the observed kinematics are the presence of strong outflows or, alternatively, a merger.

In support of the outflow picture, \cite{Finkelstein2009} and \cite{Dessauges2010} reported that ISM absorption lines in the rest-frame UV spectrum of J0022 are blueshifted with respect to the stellar photospheric features of about $120-160$ km/s. Moreover,  \cite{Shirazi2014}, modelling the H$\beta$ spatially resolved emission line velocity map in the source plane, showed that the velocity field cannot be fitted by a simple rotating disk and that more complex kinematics, possibly associated with outflows, are present.

\begin{figure}[t]
\centering
\includegraphics[width=9.5cm,trim=90 350 16 60,clip]{{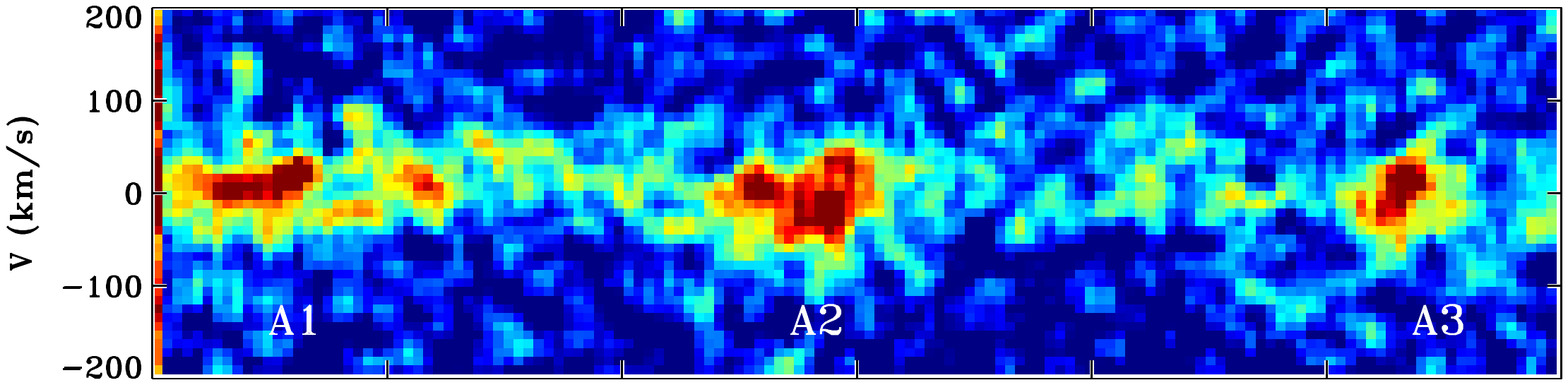}}
\includegraphics[width=9.5cm,trim=10 10 10 56,clip]{{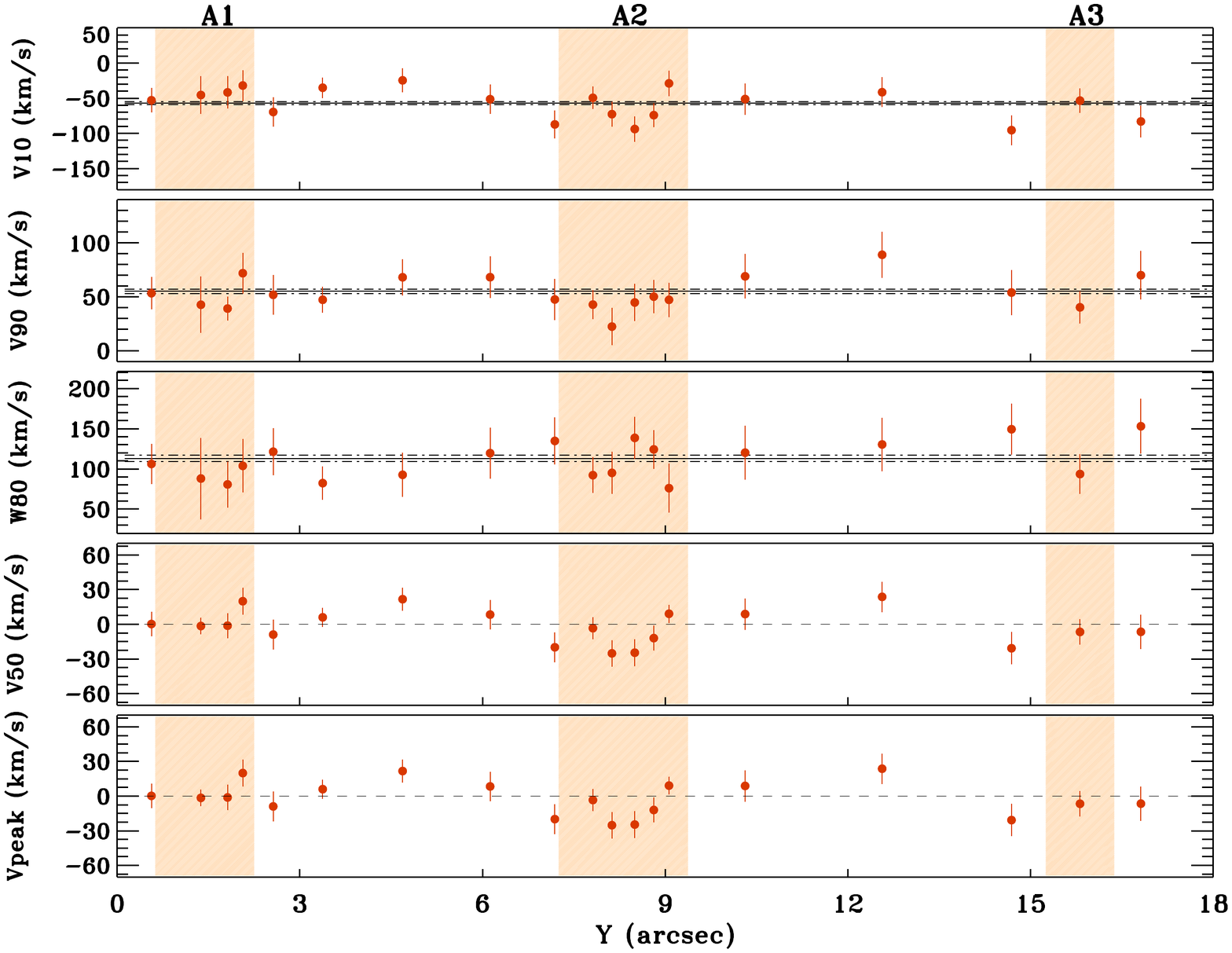}}
\caption{\small {\it Top panel:}  Zoom of the J1110 2D spectrum around the H$\alpha$ line. {\it Lower panels:} J1110 non-parametric velocity measurements as a function of the position along the dispersion axis from which 1D spectra have been extracted. Red symbols refer to H$\alpha$ velocity measurements.   See Fig. \ref{nonparametricJ1038} for a more detailed description of the figure.
}
\label{nonparametricJ1110}
\end{figure}

\subsection{J1110}
J1110+6459 (J1110 hereinafter) is a giant arc at z $= 2.48$ lensed by a galaxy cluster at redshift 0.66. It was discovered as part of the CASSOWARY survey (\citealt{Stark2013}) and is composed of three main images (A1, A2 and A3; Fig. \ref{images})  covering $\sim 17''$ on the sky. Archival HST/WFC3 (UVIS and IR) images reveal the presence of bright emission knots within the three overlapping images. \cite{Johnson2017} identified the same set of individual clumps within the three images and confirmed that they are associated with the same emitting region in the source plane. 

In Figure \ref{integratedspectra2}, we show the LBT/ARGOS integrated spectrum over the entire curved slit with the best-fit model superimposed. The H$\alpha$ line is well modelled with a single narrow Gaussian profile. However, the 2D spectrum shows more complex kinematics, with different knots elongated along the velocity axis. As for J1038, we derived a kinematic map of the H$\alpha$ emission line analysing 1D spectra extracted along the Y axis. Figure \ref{nonparametricJ1110} shows that a velocity gradient could be present in the central image, A2. It has the highest magnification and shows the most resolved structures (\citealt{Johnson2017});  however, a higher S/N is required to improve the quality of the kinematic analysis and confirm the presence of distinct kinematic structures within the A2 image. 
Kinematic variations in the non-parametric measurements are much smaller than the ones observed in the targets previously presented (i.e. J1038, J1958 and J0022); on the basis of this consideration, we may exclude a significant contribution of mergers or outflows in the observed kinematics. Instead, the kinematic gradients observed in the ARGOS spectrum may trace gravitational motions in the galaxy.

Reconstructed images of this lensed galaxy in the source plane have been presented by \cite{Johnson2017} and \cite{Rigby2017}, showing clumpy star forming regions on scales as small as a few tens of parsecs. 
\begin{table*}[h]
\scriptsize
\centering
\caption{Spectroscopic best-fit results.}
\label{spectralfitresults}
\begin{tabular}{ccccccccccccc}
name &  $f_{[O III]} $  & $f_{[O III]}/f_{H\beta}$  & $f_{H\alpha}$  & $f_{[N II]}/f_{H\alpha}$ & $W80_{[O III]}$ & $W80_{H\alpha}$ & $V10_{[O III]}$  & $V10_{H\alpha}$ &$V90_{[O III]}$  & $V90_{H\alpha}$ & $Vpeak$\\
\toprule
J0143        & $725\pm 25$ & $3.1\pm 0.4$ & $680 \pm 26$ & $0.10 \pm 0.04$ & $169\pm 3$ & $143 \pm 7$ & $-29 \pm 2$ & $-49 \pm 3$ & $140 \pm 3$& $94\pm 7$ &$0\pm 2$\\ 
\toprule
J1343 & -- & -- & $69\pm 7$ & $0.08\pm 0.07$ &--&  119$\pm$10 & -- & $-57$$\pm$6 & -- & 62 $\pm$6&$0\pm 4$\\
\toprule
J1038 & $485\pm 13$ & $5.34\pm 0.35$ & $350\pm 5$ & $0.06\pm0.02$ & 182 $\pm$ 3 & $190\pm 6$  & -148 $\pm$ 7 & $-132 \pm 6$ & 57$\pm $4 & $57 \pm 3$ & $0\pm 1$\\
\hline
-- A1  & $23.4\pm 0.8$ & $3.9\pm 0.7$ & 22.5$\pm$0.6 & 0.07$\pm$0.03 & $202 \pm 5$ & $211 \pm 9$ & $-121 \pm 5$ & $-123\pm 5$ & $91\pm4$ & $87\pm 5$ &  $-60\pm 25$\\
-- A2  & $94.2\pm 1.4$ & $6.2\pm 0.6$ & 63.8$\pm 1.7$  & 0.06$\pm 0.01$ & $154 \pm 4$   & $171\pm 6$ & $-103 \pm 3$ & $-121 \pm 3$ & $50 \pm  3$  & $51\pm 6$ & $2\pm 1$\\
-- A3  & $165.1\pm 1.6$  & $6.2\pm 0.4$& 116.9$\pm 3.2$   & 0.05$\pm 0.01$ & 152 $\pm$ 3 & $163\pm 9$ & $-103 \pm$ 2 & $-113\pm 8$ & 51 $\pm $ 2& $50\pm 4$ & $-1\pm 3 $\\
-- A4  & $87.2 \pm 1.2$ & 6.0 $\pm 0.5$ & 58.1$\pm 1.3$  & 0.03$\pm 0.01$ & 165 $\pm$ 4 & $175\pm 5$ & $-116 \pm$ 1 & $-126\pm 3$ & 49 $\pm $ 3& $46\pm 5$ & $-8\pm 1$ \\
\toprule
J1958 & $14400\pm 300$ & 4.6$\pm 0.4$ & $12400\pm 500$ & $0.10\pm 0.05$ & 282 $\pm$ 7 & 269 $\pm$ 8 & $-66 \pm$ 4 & $-73 \pm $4 & 213 $\pm$ 6 & 191 $\pm$ 9&$4 \pm 4$\\ 
\hline
 -- A1   & $3500\pm 100$  & 5.3$\pm 0.7$ & $3600\pm 200$ & $0.13\pm 0.06$ &  179 $\pm$ 9 &182$\pm 11$ & $-71 \pm$ 6 & $-77\pm $7 & 101 $\pm$9 &  105 $\pm$ 11&$-5\pm 5$\\ 
 -- A2   & 5100$\pm 100$ & $3.4  \pm 0.2$ & $4300 \pm 200$ & 0.20$\pm 0.08$ & 224 $\pm$ 12 & 188 $\pm$ 7& $-82 \pm$ 3 & $-90\pm$ 5 &  145 $\pm $13 & 98 $\pm$ 7&$-3\pm 2$ \\ 
 -- B     & 2000$\pm 200$ & $4.5\pm 1.5$ & $1700 \pm 240$ & $<0.3^{**}$ & 184 $\pm$ 35   & 158 $\pm 24$ & 75 $\pm$ 30 & $99\pm $24 & 257 $\pm$ 7 &   $257 \pm$ 7& $191\pm 13$\\ 
\toprule
J1110  & -- & -- & 178$\pm$5 & $<0.16^{**}$ & -- & 113 $\pm$4 & -- & $-57$ $\pm$ 2 & -- & 55$\pm$2&$0\pm 1$\\ 
\hline
 -- A1&   -- & -- & 49$\pm$2 & --  & -- & 105 $\pm $ 5& -- & $-52$ $\pm$ 3 & -- & 53 $\pm$3& $-1\pm 2$\\
 -- A2&   --  & -- & 39$\pm2$ & -- & -- & 112 $\pm$ 6& --& $-71$ $\pm$4 & -- & 41 $\pm$4&$-15\pm 2$\\
 -- A3&   --  & -- & 6$\pm$1& --  & -- & 94 $\pm$ 28& --& $-56$ $\pm$5 & -- & 38 $\pm$5&$-8\pm 2$\\
\toprule
J0022$^*$ & --  & --  & 140$\pm$3 & 0.27$\pm$0.01 & -- & 313 $\pm$12 & -- & $-230$ $\pm$ 9 & -- & 82$\pm$9&$-62\pm 12$\\ 
\hline
 -- A1&   -- & -- & 9$\pm$2 & 0.26$\pm$0.07 & -- & 297 $\pm$ 78& --& $-197$ $\pm$77 & -- & 100 $\pm$38&$-22 \pm 16$\\
 -- A2&   -- & -- & 53$\pm$2 & 0.26$\pm$0.02 & -- & 286 $\pm$ 15& --& $-218$ $\pm$17 & -- & 68 $\pm$11&$-40 \pm 29$\\
 -- A3&   -- & -- & 36$\pm$2 & 0.25$\pm$0.02 & -- & 266 $\pm $ 17& -- & $-187$ $\pm$ 19 & -- & 78 $\pm$10&$-34 \pm 16$\\ 
\toprule  
\hline
\end{tabular}
\tablefoot{Column (1): target name and relative images in the arc-structure; Columns (2) to (5):  non-lensing-corrected fluxes (in $10^{-17}$ erg/s/cm$^2$) and flux ratios. Columns (6) to (11): non-parametric velocity measurements from [O III]5007 and H$\alpha$. Column (12): $Vpeak$ measurements.\\
$^*$: For J0022, non-parametric estimates are derived from the[NII] line instead of H$\alpha$ (see text). All velocities are not corrected for the instrument spectral resolution; by subtracting the instrumental resolution in quadrature we would obtain a difference of 20 km/s for velocities as low as 70 km/s and even smaller corrections for highest velocities (i.e. $< 10$ km/s at $v>100$ km/s).\\
$^{**}$: from a 3$\sigma$ upper limit on the undetected [N {\small II]} line. 
}
\end{table*}

\subsection{J1343}
The system SDSS J1343+4155 (J1343) was discovered by \cite{Diehl2009} and comes from the candidate SDSS merging galaxy sample (\citealt{Allam2004}). The arc is lensed by a rich galaxy cluster at z $\sim 0.42$ with a bright central luminous red galaxy. Archival HST/FWC3 (UVIS and IR) images reveal the presence of bright emission knots within a 20$''$ elongated arc (Fig. \ref{images}).

The H$\alpha$ spatial distribution in the ARGOS 2D spectrum well reproduces  the clumpy structure observed in HST images. The integrated spectrum reveals the possible presence of the [N {\small II}]$\lambda$6584 emission line (detected at $\sim 2\sigma$ level); line features are narrow and symmetric, with $W80= 119\pm 10$ km/s. The low S/N does not allow us to study spatially resolved kinematics along the Y axis. We highlight however that the H$\alpha$ emission line seems to show a shallow velocity gradient for increasing Y values along the arc: extracting two different spectra from the lower and upper parts of the 2D spectrum (see labels in Fig. \ref{integratedspectra2}), we observe a variation in the peak velocity from $-15\pm 9$ km/s to $+19\pm 10$ km/s.

IFU OSIRIS observations and a lens model of J1343 have recently been reported by \citet[CSWA28 in their work]{Leethochawalit2016}. The authors classified this lensed galaxy as a rotationally supported system, consistent with our LUCI observations.

\section{Slit illumination effects}

Slit illumination can vary during the observations because of changes in seeing and non-perfect curved-slit matching between the arc structure and the mask. While homogenous slit illumination is obvious for extended sources and sky background, changes in slit illumination can result for sources which are spatially resolved within the slit width.  These variations may be responsible for changes in the shape of the spectral lines, introducing systematic uncertainties in velocity estimates along the slit (e.g. \citealt{Campbell1915,Spronck2013}). 

As reported in Table \ref{tablog}, the slit widths of all lensed galaxies apart from J0143 are generally smaller or comparable to the spatial resolution provided by ARGOS. Therefore we do not expect to observe significant velocity shifts due to slit illumination effects.
However,  we tested the reliability of the velocity variations observed in our spectra (e.g. Figs. \ref{nonparametricJ1038} and \ref{nonparametricJ1958}), by measuring the systemic velocity of the reference star of J1958 using its stellar absorption features.  We extracted the stellar spectrum from two A frames showing significant flux variation ($\sim 25\%$) due to a shift of the star with respect to the slit centre. Then, analysing the profiles of the stellar absorption features, we measured the line centroids. Changes of the order of a few kilometres per second are found, confirming the reliability of the (much larger) velocity variations found in the individual knots of the  lensed galaxies. We also stress that the velocity variations reported for J1038 appear to be replicated over the images A2, A3, and A4 (Fig. \ref{nonparametricJ1038}) and that the same argument applies to A1 and A2 in J1958. This recurrence makes it unlikely that the observed velocity changes are due to slit illumination effects. 

The lensed galaxy J0143 has instead been observed with a curved slit
whose width is greater than the achieved spatial resolution.  Moreover, the spatially resolved kinematic analysis does not show replicated velocity signatures between the individual knots. However, the presence of similar kinematic signatures in both [O {\small III}] and H$\alpha$ 
suggests that, also for this source, slit illumination variations are not responsible for the observed kinematics along the slit.  Dedicated follow-up, high-resolution imaging observations are however required to study the structure of the bright arc of J0143.

\section{Ionisation source for the emitting gas}\label{fluxratios}
In Fig. \ref{bptargos}, we show the BaldwinPhillipsTerlevich (BPT; \citealt{Baldwin1981}) diagram for the three sources 
 for which we are able to estimate the [O {\small III}]$\lambda$5007/H$\beta$ and [N {\small II}]$\lambda$6584/H$\alpha$ flux ratios. For J1038 and J1958, we reported the flux ratios derived from the different images along the arcs and from the total integrated spectra shown in Fig. \ref{integratedspectra}; for J0143, the reported flux ratios refer to the integrated spectrum. The lines drawn in the diagram correspond to the ``maximum starburst'' curve (\citealt{Kewley2001}) and the empirical relation (\citealt{Kauffman2003}) used to separate purely SF galaxies from galaxies containing AGN. In the figure, we also report the flux ratios associated with SDSS galaxies and AGN at z $\sim 0$ (\citealt{Abazajian2009}).

Due to the large uncertainties affecting the flux ratios, we cannot
completely exclude the presence of AGN ionisation in these sources;
neither can we distinguish between possibly different ionisation
states between the different knots associated with a given galaxy
(which are all consistent within the errors). We note, however, that
similarly large [O {\small III}]$\lambda$5007/H$\beta$ ratios are
usually observed in high-redshift galaxies
(e.g. \citealt{Kashino2016,Shapley2015,Steidel2014}), and are
attributed to more extreme ISM conditions than seen in the spectra of local galaxies rather to AGN ionisation (e.g. \citealt{Kewley2013}).  Such extreme conditions could be related to galaxy interaction/merger (J1958) and outflow (J1038 and J0143) activity, as suggested by the presence of asymmetric emission line profiles in our lensed galaxies.

For J1038, we also estimated the diagnostic ratios possibly associated with outflowing gas for the integrated spectrum and the A2, A3, and A4 knots. Outflow emission has been measured integrating the flux in the $<-50$ km/s channels; that is, considering only the blue asymmetric extension of the line  profiles (the maximum redshift velocity is $\approx 50$ km/s; see Fig. \ref{nonparametricJ1038}). We note that the flux ratios associated with outflows are shifted to higher [N {\small II}]$\lambda$6584/H$\alpha$ ( with $\gtrsim 4\sigma$ significance in the integrated spectrum). This is generally observed in SF-driven (e.g. \citealt{Ho2014,Newman2012}) and AGN-driven (e.g. \citealt{McElroy2015,Perna2017b}) outflows and is interpreted as due to shock excitation (e.g. \citealt{Allen2008}),  although other scenarios have been proposed in the literature to explain this shift\footnote{For example, a metallicity enhancement can also be responsible for the shift to higher  [N {\small II}]$\lambda$6584/H$\alpha$ flux ratios (see e.g. \citealt{Perna2017b}).}. The higher flux ratios could therefore support the idea that the J1038 asymmetric line profiles are due to SF-driven outflows. The separation between systemic and outflow flux ratios cannot be derived for the other source for which we have indications of the presence of outflows (J0143) because of the presence of bad sky subtraction residuals on the blue side of the [N {\small II}]$\lambda$6584 emission line.  

The three sources observed only in the K band (i.e. with no detection of H$\beta$+[O {\small III}]) are characterised by [N {\small II}]$\lambda$6584/H$\alpha$ ratios similar to those of J1038, J1958, and J0143 (see Table \ref{spectralfitresults}). Also in this case,  therefore, we cannot completely rule out the presence of AGN in these systems. We note however that in the low-mass galaxy regime, i.e. log(M$_{\odot}$)$\lesssim10.5$, the effect of AGN contribution to the line flux is expected to be limited (e.g. \citealt{Wuyts2016,Genzel2014,Leethochawalit2016}). 

\begin{figure}[h]
\centering
\includegraphics[width=8.5cm,trim=0 0 0 0,clip]{{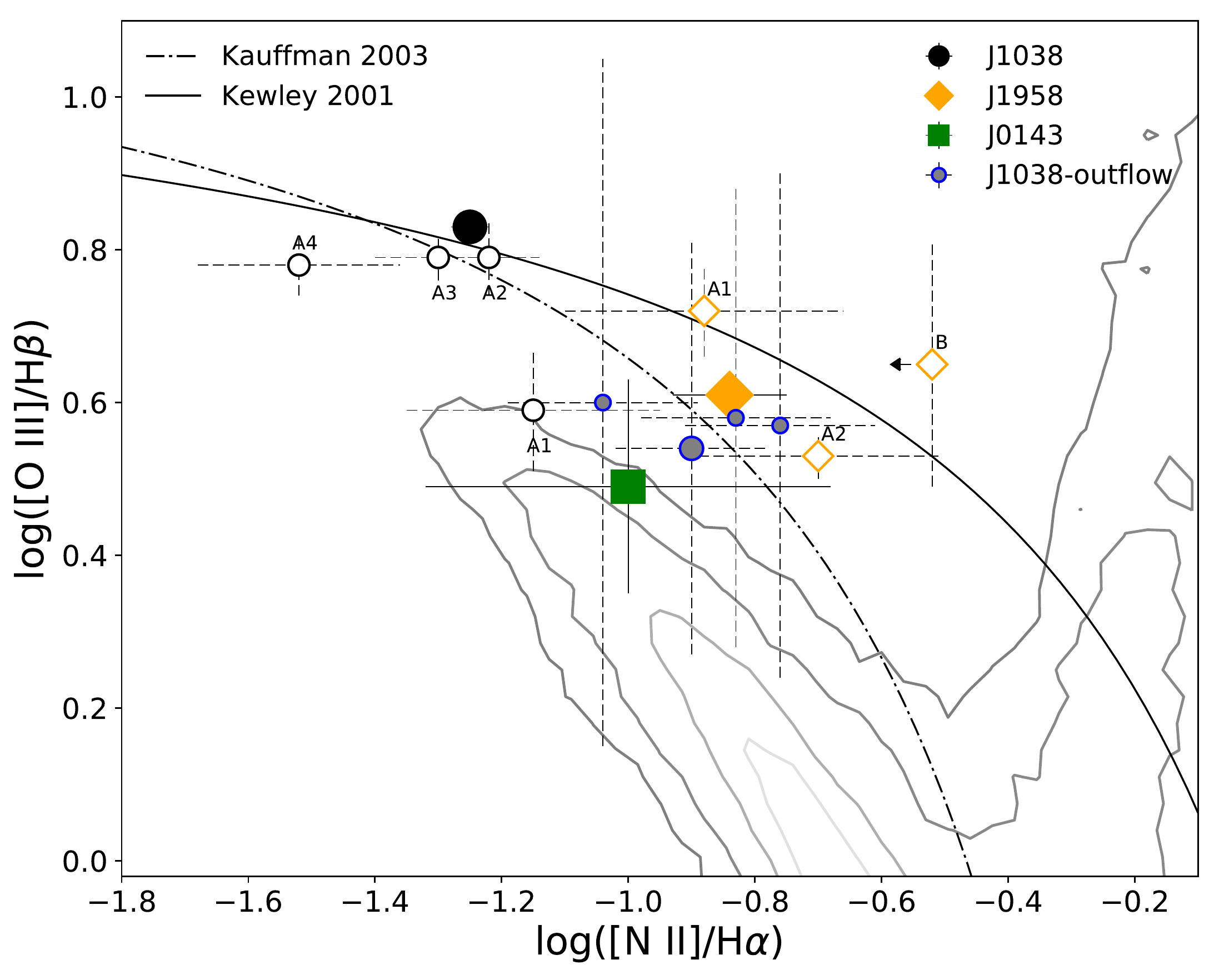}}
\caption{\small Standard diagnostic BPT diagram showing the classification scheme by \cite{Kewley2001} for our flux ratio measurements, in comparison with the SDSS (z $\sim 0$) sample (\citealt{Abazajian2009}; grey contours). The lines drawn in the figure display the curves used to separate H {\small II}-like ionised regions from galaxies containing AGN (from \citealt{Kewley2001} and \citealt{Kauffman2003}, as labelled). Black dots refer to J1038 flux ratios, orange diamonds refer to J1958 measurements, and the purple point represents J0143 line ratios. For J1038 and J1958, we report the ratios derived for each knot associated with the arc structure (A1, A2, A3 and A4 for J1038; A1, A2 and B  for J1958). For J1038, for which we have evidence for the presence of warm outflows, we report the flux ratios associated with the high-velocity gas emission for the A2, A3, and A4 images (blue/black small dots) and for the total integrated spectrum (blue/black dots; see text for details).
}
\label{bptargos}
\end{figure}

\section{Outflow mass rate and mass-loading factors}\label{outflowmassrateeta}

Outflow mass rates and mass-loading factors, which is defined as the
ratio between outflow mass rate and SFR, are important ingredients of galaxy evolution models (e.g. \citealt{OppenheimerDave2008,Hopkins2012}). In this section, we report the ionised outflow properties for the three sources for which we have good evidence of the presence of ejected material from spectroscopic analyses, that is, J1038, J0143, and J0022. We stress however that the merger scenario cannot ruled out for these targets, and that further observational efforts are required to discriminate between the two scenarios.

We derive the amount of outflowing material using Eq. (1) of \cite{Newman2012},

\begin{equation}
M_{out} = \left ( \frac {1.36\ m_H}{\gamma_{H\alpha} N_e} \times L(H\alpha) \right ) 
\end{equation}
\begin{equation*}
 = 3\times 10^8 \times \left ( \frac {L(H\alpha)} {10^{43} erg/s} \right ) \left ( \frac {N_e} {100\ cm^{-3}} \right )^{-1} M_\odot,
\end{equation*}

\noindent and, under the assumption that the gas uniformly populates a bi-conical region (e.g. \citealt{Maiolino2012}), the outflow mass rate is

\begin{equation}
\dot M_{out} = M_{out} \times 3 \frac {v_{out}}{R_{out}}  
\end{equation}

\begin{equation*}
\approx \left ( \frac {L(H\alpha)} {10^{43} erg/s} \right )  \left ( \frac {N_e} {100\ cm^{-3}} \right )^{-1}   \left ( \frac {v_{out}} {km/s} \right )   \left ( \frac {1 kpc} {R_{out}} \right ) M_\odot/yr,
\end{equation*}

\noindent
where $M_{out}$ is the mass outflow,  $\dot M_{out}$ is the mass outflow rate, $m_H$ is the atomic mass, $\gamma_{H\alpha}$ is the emissivity of the H$\alpha$ line, $N_e$ is the electron density in the outflow region, $L(H\alpha)$ is the H$\alpha$ luminosity associated with outflowing material, and $R_{out}$  and $v_{out}$ are the radial extent and velocity of the outflow. We assumed an electron density of 100 cm$^{-3}$, commonly observed in SF-driven outflowing ionised material (e.g. \citealt{Newman2012}), and an electron temperature of $T_e=10^4$ K to derive the H$\alpha$ emissivity ( $\gamma_{H\alpha}=3.56\times 10^{-25}$ erg/s cm$^3$; \citealt{Newman2012}). 

For J1038, we derived an outflow mass of $M_{out} \sim 2.5\times 10^7$ M$_{\odot}$, 
using the H$\alpha$ flux coming from the high-velocity gas in the A2 image (see Sect. \ref{fluxratios}), the one to which lens magnification correction can be applied (\citealt{Jones2013}). 
From Eq. (2), we derived an outflow mass rate $\dot M_{out} \sim 3$ M$_{\odot}$/yr, assuming a galaxy-wide extension of the ejected material, $R_{out} \sim 3$ kpc, and an outflow velocity of $v_{out}= -130$ km/s (i.e. assuming that the $V10$ non-parametric estimate is representative of the outflow velocity; e.g. \citealt{CanoDiaz2012}). We note that the outflow size $R_{out}$ cannot be derived with the available information; we therefore adopted an outflow radius corresponding to the average outflow size measured in star forming galaxies at z $\sim 2$, $R_{out} = 3$ kpc (from SINS, zC-SINF and KMOS$^{3D}$ IFU spectroscopic surveys; \citealt{Schreiber2014,Genzel2014,Newman2012}). We caution however that an even larger extension could be present in this system: \cite{Jones2013} reported the presence of small substructures associated with blueshifted gas detected in H$\alpha$ with low S/N at distances of 3-6 Kpc from the central galaxy in the reconstructed source plane. 
Finally, we derived the mass loading factor, $\eta= \dot M_{out}/SFR\approx 0.1$, using a SFR of 40 M$_{\odot}$/yr (Table \ref{tablog}\footnote{Similar SFR is obtained by \cite{Jones2013}, from the SED fitting.}).

Using similar arguments, we derived the outflow mass,  $M_{out}=3\times 10^7$ M$_\odot$, the outflow mass rate, $\dot M_{out}\sim 3$ M$_\odot$/yr, and the mass loading factor $\eta \sim 0.04$ for the J0143 lensed galaxy. For this system, we adopted the same assumptions regarding outflow geometry and plasma properties. The H$\alpha$ luminosity used to compute the outflow energetics was derived from the high-velocity ($v>50$ km/s) emission of the integrated spectrum, and an outflow size of 3 kpc was assumed following the arguments mentioned above. Due to the presence of extended red wings in the emission lines, we assumed that the outflow velocity corresponds to the measured redshifted maximum velocity in proximity of the A1 region, $v_{out}=V90 = 100$ km/s.

Finally, adopting similar arguments and a higher electron density ($N_e=630$ cm$^{-3}$; \citealt{Shirazi2014}), we derived the outflow energetics of J0022 from the A2 image information: $M_{out}=1\times 10^7$ M$_\odot$, $\dot M_{out}\sim 3$ M$_\odot$/yr, and $\eta \sim 0.01$. The presence of a strong atmospheric sky-line on the blue side of the H$\alpha$ line led us to use the [N {\small II}] line instead to constrain the fraction of flux associated with outflowing gas. Assuming the same partition
between perturbed and unperturbed fluxes for both nitrogen and H$\alpha$ emission, the calculated outflowing gas amounts to 35\% of the total flux of the H$\alpha$. This fraction is also consistent with the one obtained from H$\beta$ line by \cite{Shirazi2014}. Outflow size of $2$ kpc (from blueshifted emission extension in the reconstructed source plane; \citealt{Shirazi2014}) and velocity of 108 km/s have also been assumed.

The outflow kinetic power, $\dot E_{out}= 0.5 \dot M_{out} v_{out}^2$, is $8\times 10^{37}$ erg/s, $1\times 10^{40}$ erg/s and $1\times 10^{39}$ erg/s for J1038, J0143, and J0022, respectively. These energetics are $<< 0.1\%$ of the kinetic output expected from stellar processes (following \citealt{Veilleux2005}, $\dot E_{out} (SF)\approx 10^{42-43.5}$ erg/s); the observed outflows can therefore
be associated with SF-driven winds. All derived outflow properties are reported in Table \ref{outflowproperties}. We stress here that these outflow properties shall be considered as order of magnitude estimates, as the uncertainties can be as high as one order of magnitude themselves (e.g. \citealt{Perna2015b}).

Keeping in mind the large uncertainties and the assumptions adopted to derive outflow properties, the mass loading factors derived for these sources ($\eta \sim 0.01-0.10$) are compatible with those observed in both local luminous infrared galaxies (\citealt{Arribas2014,Chisholm2017,Cresci2017}) and high-z SFGs (\citealt{Genzel2014,Newman2012}). We also note that the difference of greater than or approximately equal to one order of magnitude in SFR and stellar mass of these systems is not reflected in significant variations in mass loading factor as  some theoretical models alternatively predict (e.g. \citealt{Muratov2015,Hayward2016}).

\begin{table}[h]
\footnotesize
\centering
\caption{Approximate outflow properties.}
\label{outflowproperties}
\begin{tabular}{lcccc}
source & $M_{out}$ (M$_\odot$) & $\dot M_{out}$ (M$_\odot$/yr) & $\dot E_{out}$ (erg/s) & $\eta$\\
\toprule
J1038 & $2.5\times 10^7$ & 3 & $8\times 10^{37}$ &  0.10\\
J0143 & $3.0\times 10^7$ & 3 & $1\times 10^{40}$ &  0.04\\
J0022 & $1.0\times 10^7$ & 6 & $1\times 10^{39}$ &  0.01\\
\hline
\end{tabular}
\end{table}

\section{Gas-phase metallicity}\label{gasmetallicity}

In this section, we present the gas-phase metallicity estimates for our sample of lensed galaxies, as we have shown that their emission lines originate predominantly from star forming regions, with negligible contribution from AGN or shocks. 
To estimate the metallicities, we used the N2 $=$ [N {\small II}]$\lambda$6584/H$\alpha$  and, when available, the R3 $=$ [O {\small III}]$\lambda$5007/H$\beta$ diagnostic ratios, assuming the $T_e$-based calibrations from \citet{Curti2017}.
When both line ratios are available, they are jointly used to put tighter constraints on metallicity, by minimising the $\chi^2$ in the N2 and R3 calibrations simultaneously while taking into account both measurement errors and dispersion of each calibration.
We run an MCMC to generate a posterior probability distribution for log(O/H), whose median and 1$\sigma$  intervals are then assumed as the inferred metallicity and its uncertainties, respectively.
Since both indicators involve lines close in wavelength, they are unaffected by nebular attenuation. We note here that this approach provides more accurate metallicity estimates than using the direct combination of the two line ratios as a single diagnostic [i.e. O3N2 $=$ ([O {\small III}]$\lambda$5007/H$\beta$ )/([N {\small III}]$\lambda$6584/H$\alpha$)] in the majority of cases, although the values inferred with the two methods are fully consistent with each other within their uncertainties. This procedure also permits one to discriminate between the two metallicity solutions for the [O {\small III}]$\lambda$5007/H$\beta$ ratio (see Fig. 9 in \citealt{Curti2017}).  
 In Table \ref{metallicityresults} we report the metallicities derived from the flux ratios previously computed both for the integrated spectra and for the different images along the arcs for all our targets\footnote{R3 metallicity estimates are not reported in Table \ref{metallicityresults} because of their lower reliability compared with the values obtained from the $\chi^2$ approach.}. 
The metallicities inferred from the combination of N2 and R3 diagnostics are in good agreement with those derived from the N2 only.

Previous metallicity estimates are reported in the literature for  J0022 (\citealt{Dessauges2011,Finkelstein2009}), J1343, and J1958 (\citealt{Leethochawalit2016}), where they are computed following the \citet{Pettini2004} calibration of the N2 indicator.
We can therefore infer the [N {\small II}]/H$\alpha$ ratio by reversing the \cite{Pettini2004} formula and compare these measurements with those derived from our LBT/ARGOS spectra. Previous metallicity estimates are also available for J1038; the comparison between our results and those presented by \citet{Jones2013} is discussed in the following section.

For J1343 and J1958, \citet{Leethochawalit2016} provides integrated [N {\small II}]/H$\alpha$ fluxes from OSIRIS observations (see Table 4 of their paper, where J1343 is named after cswa28, and J1958 is named after cswa128); 
these values are in good agreement with those reported in this work (see Table 2).
Moreover, assuming the central metallicity (reported in Table 4 of \citealt{Leethochawalit2016} ) as a representative value for these galaxies, the inferred N2 ratios for J1343 and J1958 are 0.100 $\pm$ 0.019 and 0.207$\pm $ 0.003, respectively; re-deriving then the metallicities according to our method and the C17 calibrations provides consistent estimates ($8.42\pm 0.17$ and $8.58\pm 0.16$ for J1343 and J1958 respectively) compared to those presented in this work (see the N2-column of Table 4).
These results also suggest that slit-loss effects in our spectra are not strongly affecting
line-ratio measurements in these sources if compared to IFU observations. 

For J0022, our integrated N2 ratio associated to the A2 image ($0.26 \pm 0.02$) is more than a factor of two higher than that reported by \citet[][i.e. $0.11 \pm 0.03$]{Dessauges2011}, while being fully consistent with the value reported by \citet[][i.e. $0.27\pm 0.03$]{Finkelstein2009}. 
Therefore, it follows that the oxygen abundance estimated from \citet{Dessauges2011} is underestimated ($8.43 \pm 0.19$ when reported to the abundance scale defined by the C17 calibrations) compared both to the value 
inferred from \citet[][$8.64 \pm 0.20$]{Finkelstein2009} and the LBT/ ARGOS spectrum one ($8.63 \pm 0.16$). 
 This can be reasonably explained by the low S/N in the \citet{Dessauges2011} X-Shooter spectrum (e.g. [NII] lines are detected at $2-3\sigma$ level).

 The metallicity estimates we derived for the sample of lensed
 galaxies are in agreement with those generally reported for galaxy
 systems at z$\sim 2$
 (e.g. \citealt{Erb2006,Mannucci2010,Cresci2010,Wuyts2016}). In
 Fig. \ref{MstarZargos}, we show the mass-metallicity relation (MZR)
 combining our results with those derived from stacked spectra
 analyses and reported by \cite{Erb2006}, \cite{Wuyts2014} and \cite{Steidel2014}. For consistency, all the metallicity measurements from previous works have been re-derived according to the \citet{Curti2017} calibration of the N2 indicator. 
The redshift evolution of the mass-metallicity relation is highlighted in the figure by comparing the best-fit MZR obtained by \citet{Andrews2013} from $T_e$ measurements on stacked spectra (in narrow mass bins) of SDSS galaxies. In addition, the 1$\sigma$ metallicity dispersion of the median MZR in bins of stellar mass, derived on local galaxies with the \citet{Curti2017} calibrations, is enclosed by the grey contours (see Curti et al., in prep).

\begin{figure}[h]
\centering
\includegraphics[width=8.5cm,trim=0 0 0 0,clip]{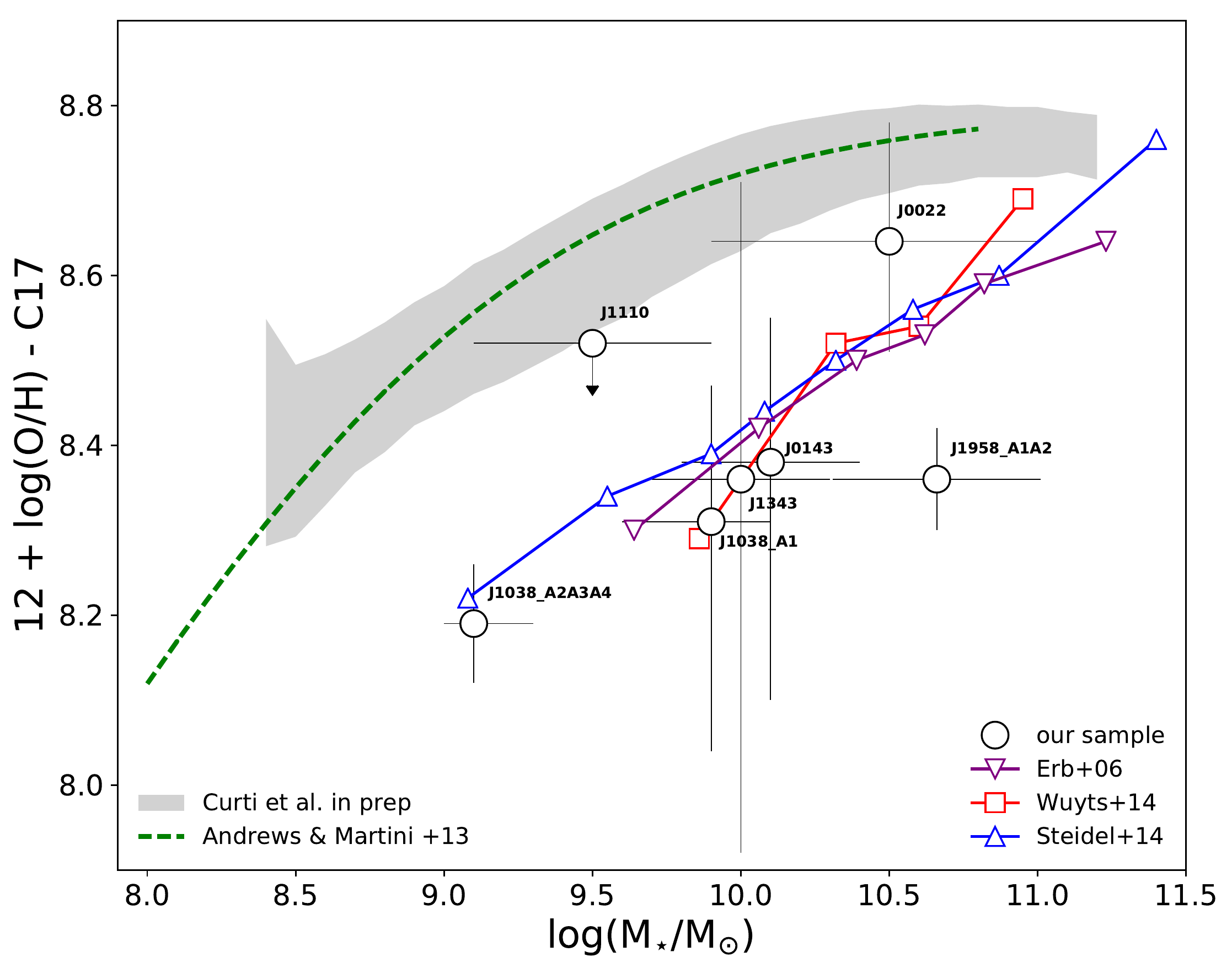}
\caption{\small 
Mass-metallicity relation (MZR) for our sample. The metallicities are derived exploiting the \citealt{Curti2017} (C17) calibration of the N2 indicator (for J1038, J1958 and J0143 we reported in the N2+R3 values, affected by smaller uncertainties; see Table \ref{metallicityresults}). Previous N2-based metallicities from the literature are plotted with different colours/symbols as reported in the legend. The grey shaded area encompass the 1$\sigma$ dispersion of the local median MZR in bins of stellar mass, as derived from the C17 calibrations for SDSS galaxies (Curti et al., in prep.).
The $T_e$-based local MZR derived from staked spectra in bins of stellar mass by \cite{Andrews2013} is also shown as the dashed green line.
We applied corrections to the stellar mass scale to account for the different IMFs adopted by previous works.
For J1038 and J1958, integrated spectra over the knots associated with an individual emitting source have been used to maximise the S/N in the faint [N {\small II}] emission lines (see Sects. \ref{j1038} and \ref{j1958} for details). Following \cite{Jones2013}, for the J1038 system we distinguished between the A1 image and the other knots, possibly associated with two interacting galaxies of $10^{9.9}$ M$_\odot$ and $\sim 10^{9.1}$ M$_\odot$, respectively. 
}
\label{MstarZargos}
\end{figure}
\begin{figure}[h]
\centering
\includegraphics[width=9.cm,trim=0 0 0 0,clip]{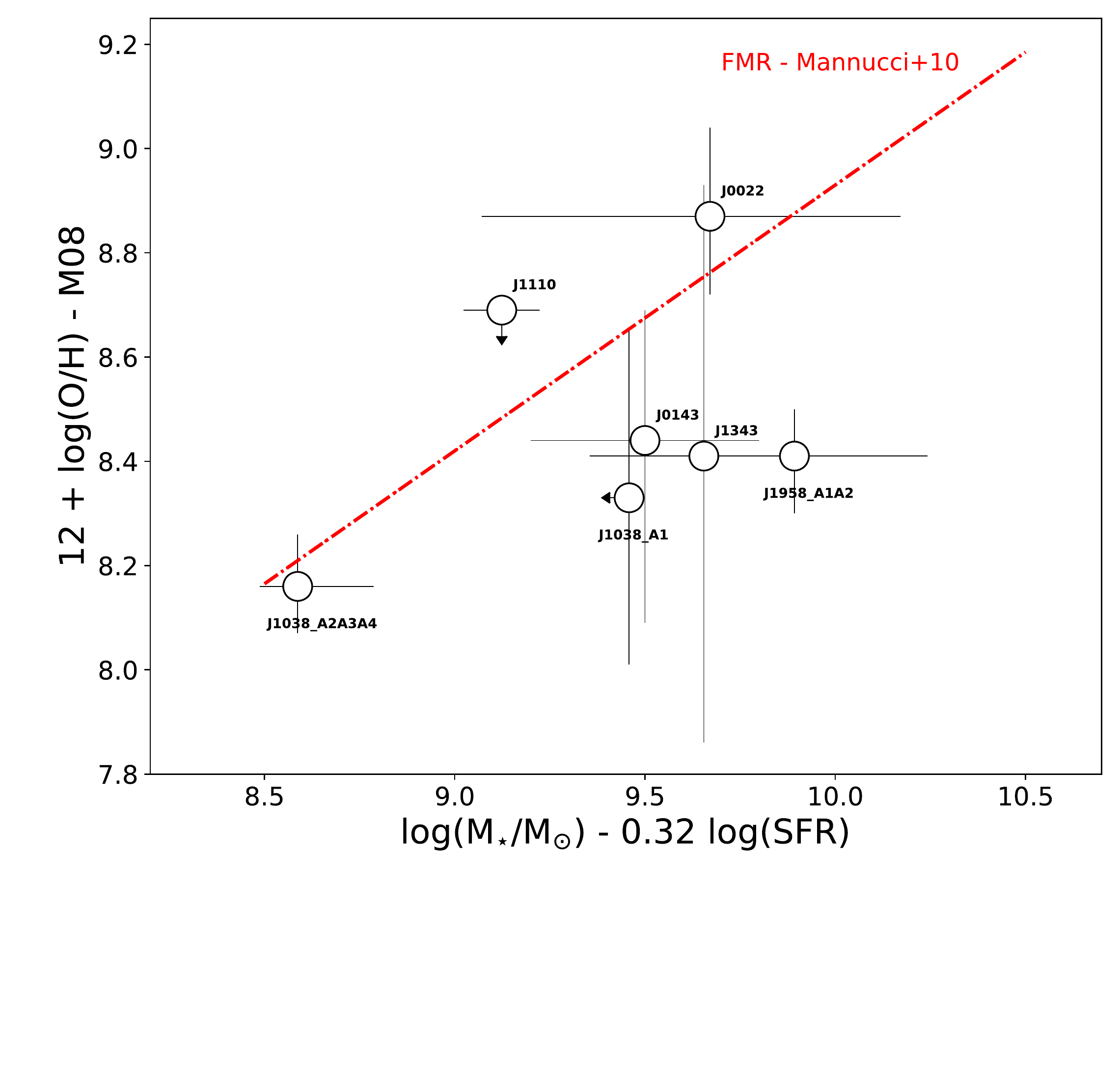}
\caption{\small 
Oxygen abundance as a function of $\mu$32, according to the \cite{Mannucci2010} 2D projection of minimum scatter of the local FMR. Metallicities are derived, for consistency on the absolute abundance scale, with the \citet[][M08]{Maiolino2008} calibrations.
}
\label{MuZargos}
\end{figure}

Finally, Fig. \ref{MuZargos}  shows the 2D projection of the fundamental metallicity relation proposed by \cite{Mannucci2010}, 12+ log(O/H) versus $\mu_{0.32}=$ log(M$_*$)$-0.32\times$ log($SFR$). In this plot, metallicities are computed exploiting the \cite{Maiolino2008} calibration of the N2 diagnostic (last column of Table \ref{metallicityresults}), 
for consistency on the absolute abundance scale with \cite{Mannucci2010}. 
Our results are in fair agreement, within the uncertainties, with the absence of any redshift evolution of the relation at these epochs (see e.g. \citealt{Belli2013, Christensen2012,Cresci2012}), although some of the points appear to lie more than 0.1 dex below the metallicity value predicted by the local FMR. In particular, J1958 is the target with the most significant deviation from the FMR; this is however consistent with recent results showing that merger galaxies increase the scatter in the FMR (e.g. \citealt{Wang2017}).

We note that most of the host galaxy properties reported in this paper have been collected from the literature, as summarised in Table \ref{tablog}. The SFRs of J1038 and J0143 were instead derived from the observed H$\alpha$ fluxes, using the \citet{Kennicutt1998} relation and correcting for lensing magnification factors. For the J1958 lensed galaxy, we derived the stellar mass from SED fitting analysis, using SDSS and LBT/ARGOS Ks photometric data.

The posterior probability distribution function (PDF) for the stellar mass is obtained by comparing the fluxes in the six available passbands (ugriz and Ks) with the corresponding synthetic fluxes derived from a subset of the suite of 500,000 stellar population synthesis models realised by \cite{Zibetti2017}, restricted to match the age constraint imposed by the redshift of the lensed galaxy. The mass corresponding to the median of the PDF is adopted as our best estimate, while the uncertainty is given by half of the 16th-84th percentile range (we note that the magnification factor uncertainty is not taken into account and that the error reported in Table \ref{tablog} should be taken as a lower limit).

\begin{table*}[h]
\footnotesize
\centering
\caption{Metallicity}
\label{metallicityresults}
\begin{tabular}{lr|cc|cc}
\multicolumn{2}{c}{name} &  \multicolumn{2}{c}{12+log(O/H) [C17]} & \multicolumn{2}{c}{12+log(O/H) [M08]}\\
       &   &  (from N2) & (from N2+R3) & (from N2) & (from N2+R3)\\
\toprule
J1343 &  total        & $8.36_{-0.44}^{+0.35}$ &  -- & $8.41_{-0.55}^{+0.52}$&--\\
\toprule
J0143 &   total          & $8.42_{-0.36}^{+0.35}$ & $8.38_{-0.28}^{+0.17} $ & $8.50_{-0.51}^{+0.48}$&$8.43_{-0.35}^{+0.25}$ \\
\toprule
 J1038 & total$^*$   & $8.28_{-0.15}^{+0.14}$ & $8.19_{-0.07}^{+0.07}$ & $8.31_{-0.18}^{+0.17}$&$8.16_{-0.09}^{+0.10}$\\
          &A1        & $8.34_{-0.32}^{+0.24}$ & $8.31_{-0.27}^{+0.16}$ & $8.36_{-0.38}^{+0.39}$&$8.33_{-32}^{+24}$\\
          &A2         & $8.29_{-0.21}^{+0.19}$ &$8.23_{-0.13}^{+0.12}$ & $8.33_{-0.26}^{+0.24}$&$8.20_{-0.16}^{+0.16}$\\
          &A3         & $8.26_{-0.24}^{+0.19}$ &$8.17_{-0.16}^{+0.13}$ & $8.29_{-0.29}^{+0.28}$&$8.15_{-0.19}^{+0.15}$\\
\hline
          &A3 (low) & $7.99_{-0.25}^{+0.23}$ &$8.12_{-0.30}^{+0.19}$ & $7.90_{-0.31}^{+0.28}$& $7.96_{-0.36}^{+0.30}$\\
 &A3 (centre)     & $8.31_{-0.16}^{+0.14}$ & $8.28_{-0.07}^{+0.05}$ & $8.35_{-0.20}^{+0.19}$& $8.26_{-0.10}^{+0.09}$\\
 &A3 (up)         & $8.12_{-0.19}^{+0.17}$ & $8.13_{-0.19}^{+0.09}$ & $8.11_{-0.21}^{+0.20}$& $8.08_{-0.17}^{+0.11}$\\
\hline
          &A4         & $8.10_{-0.31}^{+0.25}$ &$8.08_{-0.26}^{+0.19}$ & $8.06_{-0.42}^{+0.38}$& $8.02_{-0.29}^{+0.23}$\\
\toprule
J1958 & total$^{*}$     & $8.49_{-0.19}^{+0.19} $ & $8.36_{-0.06}^{+0.06}$ & $8.65_{-0.25}^{+0.27}$&$8.41_{-0.11}^{+0.09}$\\
          & A1        & $8.48_{-0.27}^{+0.27} $ &$8.29_{-0.17}^{+0.13}$ & $8.59_{-0.38}^{+0.42}$& $8.28_{-0.21}^{+0.21}$\\
          & A2        & $8.56_{-0.25}^{+0.32}$ &$8.39_{-0.10}^{+0.08}$ & $8.76_{-0.36}^{+0.26}$& $8.46_{-0.14}^{+0.13}$\\
          & B          & $<8.87$             & $8.16_{-0.16}^{+0.34}$ & $<9.00$&$8.22_{-0.32}^{+0.39}$\\
\toprule
J1110 &   total       & $< 8.52$& -- & $<8.69$&--\\
\toprule
J0022 & total    & $8.64_{-0.13}^{+0.14}$ & -- & $8.87_{-0.15}^{+0.17}$&--\\
          & A1       & $8.62_{-0.14}^{+0.17}$ & -- & $8.83_{-0.18}^{+0.20}$&--\\
          & A2       & $8.63_{-0.14}^{+0.17}$ & -- & $8.86_{-0.18}^{+0.17}$&--\\
          & A3       & $8.61_{-0.21}^{+0.28}$ & -- & $8.87_{-0.31}^{+0.16}$&--\\
\toprule
\toprule
\end{tabular}
\tablefoot{Gas-phase metallicity estimates from the N2 and R3+N2 diagnostics with the calibration from \citet[C17; columns 3 and 4]{Curti2017} and \citet[M08; columns 5 and 6]{Maiolino2008}. 
Both the intrinsic scatter of the calibrations and the flux ratio uncertainties are taken into account in computing the total uncertainties. For J1038, the R3 ratios are derived rescaling the H$\alpha$ flux (see the text for details).\\
$^*$: the metallicity measurements were derived from the spectrum obtained integrating the flux over the A2, A3 and A4 knots, for J1038 (with log([OIII]/H$\beta$) $=0.83\pm 0.01$ and log([NII]/H$\alpha$) $=-1.25\pm 0.03$), and over the A1 and A2 knots, for J1958 (log([OIII]/H$\beta$) $=0.61\pm 0.01$ and log([NII]/H$\alpha$) $=-0.84\pm 0.09$), which can be associated with unique emitting regions in the source plane.
}
\end{table*}

\subsection{Metallicity gradient in J1038}
The large uncertainties in metallicity estimates due to the low S/N in the detected emission lines and the intrinsic scatter in the metallicity calibrations do not allow us to unveil the possible presence of different gas-phase metallicity conditions in spatially resolved regions and between the different arc images.  According to the measured errors, however, available data constrain spatial metallicity variations to values lower than $\approx 0.15$ dex, consistent with recent observations and model predictions (see e.g. \citealt{Wang2017} and references therein). Greater time investment is required to spatially resolve the metallicity distribution in these systems.

J1038 is the source with the highest S/N and has been observed in both  [O {\small III}]+H$\beta$ and [N {\small II}]+H$\alpha$ regions. If we consider the brightest and best-resolved image of the source, A3, and extract individual spectra requiring a minimum S/N of 2 for the [N {\small II}]$\lambda$6584 detection, we can obtain three spectra associated with the upper and lower blobs  (see Fig. \ref{integratedspectra}) and the fainter central region. We fitted these spectra using the same constraints introduced in Sect. \ref{j1038} and derived the line ratios to estimate the metallicity within different regions in the galaxy. Given that the Balmer decrement H$\alpha$/H$\beta$ obtained from the integrated spectrum is consistent with the absence of dust-extinction, and that the H$\beta$ line is affected by strong sky line residuals at shorter wavelengths, we used the case B ratio of 2.86 (\citealt{Gaskell1984}) to derive  [O {\small III}]$\lambda$5007/H$\beta$ ratios rescaling the H$\alpha$ fluxes\footnote{The metallicity estimates derived using the measured log([O III]/H$\beta$) are consistent with those obtained from the `corrected' flux ratios within the errors, although associated with larger scatter.}. 

Our metallicity measurements for this source are consistent with the ones reported by \cite{Jones2013} for the central regions in the reconstructed source image plane.  Indeed, N2 and R3 emission line ratios from our ARGOS integrated spectra are consistent within 1$\sigma$ with those inferred from the total observed fluxes by the OSIRIS spectrograph for the same source. However, we also note that  despite the larger S/N in our spectra with respect to those reported by \cite{Jones2013}, we do not detect any trend in the metallicity of J1038. Moreover, even using the same calibrations (from \citealt{Maiolino2008}), our estimates do not show the large range of metallicities found when analysing Keck spectra, with 12+log(O/H) between 8.1 and 9.0 (see the last column  of Table \ref{metallicityresults}). A possible cause lies in the low S/N of the faint emission lines used: for example, the [N {\small II}]$\lambda$6584 line is detected in A2 with a significance of $\sim 2\sigma$ ($1\sigma$) in the LBT/ARGOS (Keck) integrated spectrum.

For J1038, we derived an upper limit on the metallicity gradient of $0.12$ dex/kpc ($0.09$ dex/kpc), following the \citealt{Maiolino2008} (\citealt{Curti2017}) calibration of the N2 indicator and considering 2 kpc as the minimum distance sampled in the source plane by the three spectra extracted from A3 (see Fig. 3 of \citealt{Jones2013}). This metallicity gradient upper limit is lower than but still compatible within the errors with the one reported by Jones et al. ($<0.12$ dex/kpc vs. $0.15\pm 0.07$ dex/kpc, following the same calibration).

\section{Summary}

We present spatially resolved kinematics and gas-phase metallicity measurements for six gravitationally lensed galaxies at z $\sim 2$ based on the analysis of rest-frame optical emission lines. The entire extension of the arc-like geometries was observed with the LBT laser-assisted adaptive optics system ARGOS. Near infrared spectra with high spectral ($\sim 50$ km/s) and spatial ($0.4''$) resolution have been obtained to cover the H$\alpha$+[N {\small II}] complex redshifted in the H- and K bands. For three sources, we also covered the H$\beta$+[O {\small III}] lines.  

We combined ionised-gas spatially resolved spectroscopic analysis with
available spectroscopic information and lens modelling to constrain
the kinematic and chemical properties of our targets. In particular, we find the following key results.

We find strongly disturbed kinematics in the ISM ionised component of the lensed galaxy J1038. Broad and asymmetric line profiles, with $W80$ as high as $\sim 150\div300$ km/s and $V10\approx -100\div-200$ km/s, have been detected over the entire extension of the arc-like structure. In \cite{Jones2013} and \cite{Leethochawalit2016}, this target has been associated with an undergoing merger, with the A1 and A2 knots associated with two companions. Our analysis shows that the A1 image can be associated with a different emitting region in the source plane. We also suggest that the disturbed kinematics and diagnostic line ratios we derived for the other images can be related to the presence of SF-driven outflow in this system.

The spatially resolved non-parametric velocity measurements along the arc of J1958 reveal the presence of two distinct, spatially separated kinematic components, suggesting the presence of two approaching galaxies in this system. The source plane reconstruction presented by \cite{Leethochawalit2016} confirms our results.

Complex kinematics have also been observed in the H$\alpha$-[N {\small II}] system of J0022.  Our analysis revealed the presence of blueshifted, possibly outflowing, gas in the region between the two brightest images, A1 and A2, confirming the results previously reported in literature (\citealt{Shirazi2014} and references therein).

The presence of redshifted emitting gas, possibly associated with receding outflowing material, has been observed in J0143.

For the three sources exhibiting disturbed kinematics reasonably
associated with ejected ionised material, we derive outflow energetics
and a mass loading factor of $\sim 0.01-0.1$, consistent with previous
studies of SFGs. 

We did not find any strong evidence for AGN activity in these systems. Assuming that the observed emission lines are mostly due to SF activity, we derived gas-phase metallicity estimates using the N2 and R3 flux ratio diagnostics. 
The average offset of our targets from the local mass-metallicity relation is  $\sim 0.25$ dex. The lensed galaxies analysed in this work are consistent, within the errors, with the local FMR.

The low-S/N spectra do not allow us to infer spatially resolved metallicity information. On the basis of our measurements, we assessed that the metallicity variations along the slit in our systems should be less than $\approx 0.15$ dex.

The analysed NIR spectra in this study represent the first scientific demonstration of the spectroscopic capabilities of the ARGOS facility in combination with LUCI. Kinematic and ionised gas properties within the individual images of lensed systems have been easily recognised thanks to the achieved high spectral and angular resolutions provided by the AO correction in combination with the use of narrow, curved-shape slits. In order to better constrain gas-phase metallicity conditions within high-z galaxies, high-S/N spectra are required. 
The simultaneous use of both LUCI instruments in binocular mode, in combination with the ARGOS facility, is already reducing the time needed to complete spectroscopic observing programs by nearly a factor of three; the full exploitation of the ARGOS correction in binocular mode will provide an even greater boost to the telescope efficiency in the near future, providing a powerful way to improve our knowledge on galaxy evolution.
\vspace{2cm}

{\small 
{\it Acknowledgments:} 
We thank the anonymous referee for his/her constructive comments on
the paper, which considerably improved the presentation of the
results. We also thank the ARGOS team and the LBTO staff for their
support during the commissioning runs. The authors thank T. Jones,
Z. Kostrzewa-Rutkowska and M. Dessauges-Zavadsky for the helpful conversation on the J1038,
J0143 and J0022 lens models. M.P. thanks R. Ca{\~n}ameras for useful discussions. M.P. acknowledges P. Sottocorona for his valuable assistance during the development of this work. G.C. acknowledges the
support by INAF/Frontiera through the ``Progetti Premiali'' funding
scheme of the Italian Ministry of Education, University, and
Research. S.Z. has been supported by the EU Marie Curie Career Integration Grant "SteMaGE" Nr. PCIG12-GA-2012-326466  (Call Identifier: FP7-PEOPLE-2012 CIG). G.C., F.M. and S.Z. have been supported by the INAF PRIN-SKA 2017 program 1.05.01.88.04. 
}

\end{document}